\documentclass{aa}
\usepackage{natbib}
\bibpunct{(}{)}{;}{a}{}{,}
\usepackage{graphicx}
\usepackage{natbib}

\newcommand{\mb}[1]{\mbox{\boldmath $#1$}}
\makeatletter
\def\gsim{\compoundrel>\over\sim}

\def\compoundrel#1\over#2%
  {\mathpalette\compoundreL{{#1}\over{#2}}}
\def\compoundreL#1#2{\compoundREL#1#2}
\def\compoundREL#1#2\over#3%
  {\mathrel{\vcenter{\hbox{$\m@th \buildrel{#1#2}\over{#1#3}$}}}}
\makeatother

\begin{document}

\title{Exploring the relativistic regime with Newtonian hydrodynamics: \\
  An improved effective gravitational potential for supernova simulations}

\author{A.~Marek \and H.~Dimmelmeier \and H.-Th.~Janka \and E.~M\"uller \and R.~Buras}

\offprints{H.-Th.~Janka, \\ \email{thj@mpa-garching.mpg.de}}

\institute{Max-Planck-Institut f\"ur Astrophysik,
  Karl-Schwarzschild-Stra{\ss}e 1, 85741 Garching, Germany}

\date{Received date / Accepted date}

%%%%%%%%%%%%%%%%%%%%%%%%%%%%%%%%%%%%%%%%%%%%%%%%%%%%%%%%%
%%%%%%%%%%%%%%%%%%%%%%%%%%%%%%%%%%%%%%%%%%%%%%%%%%%%%%%%%
% ABSTRACT

\abstract{We investigate the possibility approximating relativistic
  effects in hydrodynamical simulations of stellar core collapse and
  post-bounce evolution by using a modified gravitational potential in
  an otherwise standard Newtonian hydrodynamic code. Different
  modifications of a previously introduced effective relativistic
  potential are discussed. Corresponding hydrostatic
  solutions are compared with solutions of the TOV equations, and
  hydrodynamic simulations with two different codes are compared with
  fully relativistic results. One code is applied for one- and
  two-dimensional calculations with a simple equation of state, and
  employs either the modified effective relativistic potential in a
  Newtonian framework or solves the general relativistic field
  equations under the assumption of the conformal flatness condition
  (CFC) for the three-metric. The second code allows for full-scale
  supernova runs including a microphysical equation of state and
  neutrino transport based on the solution of the Boltzmann equation
  and its moments equations. We present prescriptions for the
  effective relativistic potential for self-gravitating fluids to be
  used in Newtonian codes, which produce excellent agreement with
  fully relativistic solutions in spherical symmetry, leading to
  significant improvements compared to previously published
  approximations. Moreover, they also approximate qualitatively well
  relativistic solutions for models with rotation.

  \keywords{gravitation -- hydrodynamics -- methods: numerical --
    relativity -- stars:supernovae:general}
}

\authorrunning{A.~Marek et al.}
\titlerunning{An improved effective potential for supernova
  simulations}

\maketitle

%%%%%%%%%%%%%%%%%%%%%%%%%%%%%%%%%%%%%%%%%%%%%%%%%%%%%%%%%%%%%%%%%
%%%%%%%%%%%%%%%%%%%%%%%%%%%%%%%%%%%%%%%%%%%%%%%%%%%%%%%%%%%%%%%%%

%%%%%%%%%%%%%%%%%%%%%%%%%%%%%%%%%%%%%%%%%%%%%%%%%%%%%%%%%%%%%%%%%
%%%%%%%%%%%%%%%%%%%%%%%%%%%%%%%%%%%%%%%%%%%%%%%%%%%%%%%%%%%%%%%%%
% INTRODUCTION

\section{Introduction}
\label{sec:introduction}

It is a well known fact that gravity plays an important role during
all stages of a core collapse supernova. Gravity is the driving force
that at the end of the life of massive stars overcomes the pressure
forces and causes the collapse of the stellar core. Furthermore, the
subsequent supernova explosion results from the fact that various
processes tap the enormous amount of gravitational binding energy
released during the formation of the proto-neutron star. General
relativistic effects are important for this process and cannot be
neglected in quantitative models because of the increasing
compactness of the proto-neutron star. Therefore, it is important to
include a proper treatment of general relativity or an appropriate
approximation in the numerical codes that one uses to study core
collapse supernovae.

Recently \citet{liebendoerfer_05_a} performed a comparison of the
results obtained with the supernova simulation codes \textsc{Vertex}
and \textsc{Agile-BoltzTran} which both solve the Boltzmann transport
equation for neutrinos. The \textsc{Vertex} code
\citep[see][]{rampp_02_a} is based on the Newtonian hydrodynamics code
\textsc{Prometheus} \citep{fryxell_90_a} and utilizes a generalised
potential to approximate relativistic gravity. The
\textsc{Agile-BoltzTran} code of the Oak Ridge-Basel group
\citep{liebendoerfer_01_a, liebendoerfer_02_a, liebendoerfer_04_a,
  liebendoerfer_05_a} is a fully relativistic (1D) hydrodynamics
code. The comparison showed that both codes produce qualitatively very
similar results except for some small (but growing) quantitative
differences occurring in the late post-bounce evolution. Inspired by this
comparison we explored improvements of the effective
relativistic potential used by \citet{rampp_02_a} in order to achieve
an even better agreement than that reported by
\citet{liebendoerfer_05_a}. To this end we tested different
variants of approximations to relativistic gravity employing two
simulation codes.

On the one hand purely hydrodynamic simulations were performed
with the code \textsc{CoCoNuT} of \citet{dimmelmeier_02_a,
dimmelmeier_05_a} assuming spherical or axial symmetry. This code
optionally either uses a Newtonian (or alternatively an effective
relativistic) gravitational potential in a Newtonian treatment of the
hydrodynamic equations \emph{or} solves the general relativistic
equations of fluid dynamics and the relativistic field equations. The
latter are formulated under the assumption of the conformal flatness
condition (CFC) for the three-metric, also known as the
Isenberg--Wilson--Mathews approximation \citep{isenberg_78_a,
wilson_96_a}, which is identical to solving the exact general
relativistic equations in spherical symmetry. This allows for a direct
comparison of different effective relativistic potentials with
a fully relativistic treatment in both spherically symmetric and
axisymmetric simulations using the same code.

On the other hand, we used the computationally expensive
\textsc{Vertex} code \citep{rampp_02_a} for spherically symmetric
supernova simulations including neutrino transport and a microphysical
equation of state (EoS). The results of these calculations performed
with different effective relativistic potentials are compared with
those obtained with the general relativistic \textsc{Agile-BoltzTran}
code of the Oak Ridge-Basel group \citep[cf.][]{liebendoerfer_05_a}.

In this paper we discuss the results of our test calculations and
present improved effective relativistic potentials which are simple to
implement without modifying the Newtonian equations of hydrodynamics
in an existing code, and which approximate the results of general
relativistic simulations very well.

The paper is organised as follows: In
Sect.~\ref{sec:effective_potential} we describe the effective
relativistic potentials used in this investigation. In
Sect.~\ref{sec:numerical_methods} we briefly discuss the most
important features of the numerical codes we used for our
hydrodynamic simulations. In Sect.~\ref{sec:spherical_simulations} we
present the results obtained by applying the effective relativistic
potentials in spherically symmetric simulations of supernova core
collapse and to neutron star models, while
Sect.~\ref{sec:rotating_simulations} is devoted to a discussion of the
multi-dimensional simulations of rotational supernova core collapse.
Finally, in Sect.~\ref{sec:summary} we summarise our findings and draw
some conclusions.

Throughout the article, we use geometrised units with $ c = G = 1 $.

% END INTRODUCTION
%%%%%%%%%%%%%%%%%%%%%%%%%%%%%%%%%%%%%%%%%%%%%%%%%%%%%%%%%%%%%%%%%
%%%%%%%%%%%%%%%%%%%%%%%%%%%%%%%%%%%%%%%%%%%%%%%%%%%%%%%%%%%%%%%%%

%%%%%%%%%%%%%%%%%%%%%%%%%%%%%%%%%%%%%%%%%%%%%%%%%%%%%%%%%%%%%%%%%
%%%%%%%%%%%%%%%%%%%%%%%%%%%%%%%%%%%%%%%%%%%%%%%%%%%%%%%%%%%%%%%%%
% SECTION GRAVITATIONAL POTENTIAL

\section{Effective relativistic potential}
\label{sec:effective_potential}

Approximating the effects of general relativistic gravity in a
Newtonian hydrodynamics code may be attempted by using an effective
relativistic gravitational potential $ \Phi_\mathrm{eff} $ which
mimics the deeper gravitational well of the relativistic case. In the
following Sects.~\ref{subsec:tov_potential}
and~\ref{subsec:improved_potentials}, several of these effective
relativistic potentials will be discussed.

%%%%%%%%%%%%%%%%%%%%%%%%%%%%%%%%%%%%%%%%%%%%%%%%%%%%%%%%%%%%%%%%%
%%%%%%%%%%%%%%%%%%%%%%%%%%%%%%%%%%%%%%%%%%%%%%%%%%%%%%%%%%%%%%%%%
% SUBSECTION TOV-POTENTIAL

\subsection{TOV potential for a self-gravitating fluid}
\label{subsec:tov_potential}

For a self-gravitating fluid it is desirable that an effective
relativistic potential reproduces the solution of hydrostatic
equilibrium according to the Tolman--Oppenheimer--Volkoff (TOV)
equation. With this requirement in mind and comparing the
relativistic equation of motion \citep[cf.][]{vanriper_79_a,
  baron_89_a} with its Newtonian analogon, \citet{rampp_02_a}
rearranged the relativistic terms into an effective relativistic
potential \citep[see][for the hydrostatic, neutrino-less
case]{kippenhahn_90_a}.

Thus for spherically symmetric simulations using a Newtonian
hydrodynamics code the idea is to replace the Newtonian gravitational
potential
\begin{equation}
  \Phi (r) = - 4 \pi \int_0^\infty \!\! \mathrm{d}r' \,
  r'^2 \frac{\rho}{|r - r'|}
  \label{eq:regular_potential}
\end{equation}
by the TOV potential
\begin{eqnarray}
  \Phi_\mathrm{TOV} (r) & = & - 4 \pi \int_r^\infty
  \frac{\mathrm{d}r'}{r'^2}
  \left( \frac{m_\mathrm{TOV}}{4 \pi} + r'^3 (P + {p}_\nu) \right)
  \nonumber \\
  & & \qquad \qquad \times \frac{1}{\Gamma^2}
  \left( \frac{\rho + e + P}{\rho} \right),
  \label{eq:tov_potential}
\end{eqnarray}%
to obtain the effective relativistic potential $ \Phi_\mathrm{eff} $
as
\begin{equation}
  \Phi_\mathrm{eff} = \Phi_\mathrm{TOV}.
  \label{eq:effective_potential}
\end{equation}
Here $ \rho $ is the rest-mass density, $ e = \rho \epsilon $ is the
internal energy density with $ \epsilon $ being the specific internal
energy, and $ P $ is the gas pressure. The TOV mass is given by
\begin{equation}
  m_\mathrm{TOV} (r) = 4 \pi \int_0^r \mathrm{d}r' \, r'^2
  \! \left( \rho + e + E + \frac{v F}{\Gamma} \right),
  \label{eq:tov_mass}
\end{equation}
where $ p_\nu $, $ E $, and $ F $ are the neutrino pressure, the
neutrino energy density, and the neutrino flux, respectively
\citep{baron_89_a, rampp_02_a}.

The fluid velocity $ v $ is identified with the local radial velocity
calculated by the Newtonian code and the metric function $ \Gamma $ is
given by
\begin{equation}
  \Gamma = \sqrt{1 + v^2 - \frac{2 m_\mathrm{TOV}}{r}}.
  \label{eq:gamma_factor}
\end{equation}
The velocity-dependent terms were added for a closer match with
the general relativistic form of the equation of motion
\citep{vanriper_79_a, baron_89_a}. In the treatment of neutrino
transport general relativistic redshift and time dilation effects are
included, but for reasons of consistency with the Newtonian
hydrodynamics part of the code the distinction between coordinate and
proper radius is ignored in the relativistic transport equations
\citep[for details, see Sect.~3.7.2 of][]{rampp_02_a}. The quality of
this approach was ascertained by a comparison with fully relativistic
calculations \citep{rampp_02_a, liebendoerfer_05_a}.

In order to calculate the effective relativistic potential for
multi-dimensional flows we substitute the ``spherical contribution''
$ \overline{\!\Phi\!}\, (r) $ to the multi-dimensional Newtonian
gravitational potential
\begin{equation}
  \Phi (r, \theta, \varphi)  =
  - \int_V \!\! \mathrm{d}r' \, \mathrm{d}\theta' \,
  \mathrm{d}\varphi' \, r'^2 \sin \theta'
  \frac{\rho}{|\mb{r} - \mb{r'}|}
  \label{eq:regular_potential_multid}
\end{equation}
by the TOV potential $ \overline{\!\Phi\!}\,_\mathrm{TOV} $:
\begin{equation}
  \Phi_\mathrm{eff} = \Phi - \overline{\!\Phi\!}\, +
  \overline{\!\Phi\!}\,_\mathrm{TOV}.
  \label{eq:effective_potential_multid}
\end{equation}
Here $ \overline{\!\Phi\!}\,(r) $ and
$ \overline{\!\Phi\!}\,_\mathrm{TOV} $ are calculated according to
Eqs.~(\ref{eq:regular_potential}) and~(\ref{eq:tov_potential}),
respectively, however with the hydrodynamic quantities $ \rho $,
$ e $, $ P $, $ v $ and the neutrino quantities $ E $, $ F $,
$ p_\nu $ being replaced by their corresponding angularly averaged
values. Note that $ v $ here refers to the radial component of the
velocity, only.

% END SUBSECTION TOV-POTENTIAL
%%%%%%%%%%%%%%%%%%%%%%%%%%%%%%%%%%%%%%%%%%%%%%%%%%%%%%%%%%%%%%%%%
%%%%%%%%%%%%%%%%%%%%%%%%%%%%%%%%%%%%%%%%%%%%%%%%%%%%%%%%%%%%%%%%%

%%%%%%%%%%%%%%%%%%%%%%%%%%%%%%%%%%%%%%%%%%%%%%%%%%%%%%%%%%%%%%%%%
%%%%%%%%%%%%%%%%%%%%%%%%%%%%%%%%%%%%%%%%%%%%%%%%%%%%%%%%%%%%%%%%%
% SUBSECTION VARIANTS OF EFFECTIVE POTENTIAL

\subsection{Modifications of the TOV potential}
\label{subsec:improved_potentials}

In a recent comparison \citet{liebendoerfer_05_a} found that gravity
as described by the TOV potential in
Eq.~(\ref{eq:effective_potential}) overrates the relativistic
effects, because in combination with Newtonian kinematics it tends to
overestimate the infall velocities and to underestimate the flow
inertia in the pre-shock region. Thus, supposedly via the nonlinear
dependence of $ \Phi_\mathrm{eff} $ on $ e $ and $ P $ the compactness
of the proto-neutron star is overestimated, with this tendency
increasing at later times after core bounce. Consequently, the
neutrino luminosities and the mean energies of the emitted neutrinos
are larger than in the corresponding relativistic simulation.

In order to reduce these discrepancies -- without sacrificing the
simplicity of Newtonian dynamics -- we tested several
modifications of the TOV potential, Eqs.~(\ref{eq:tov_potential}),
which all act to weaken it. In particular, we studied the
following variations:

\begin{description}
\item[\it Case~A:] In the integrand of Eq.~(\ref{eq:tov_mass}) a
  factor $ \Gamma $, Eq.~(\ref{eq:gamma_factor}) is added. Since
  $ \Gamma < 1 $ this reduces the gravitational TOV mass used in
  the potential.
  \\ [-0.6 em]
\item[\it Case~B:] In Eq.~(\ref{eq:tov_mass}) the internal gas energy
  density and the neutrino terms are set to zero, $ e = E = F = 0 $,
  which again decreases the gravitational TOV mass.
  \\ [-0.6 em]
\item[\it Case~C:] In Eq.~(\ref{eq:tov_potential}) the internal gas
  energy is set to zero, $ e = 0 $, which directly weakens the TOV
  potential.
  \\ [-0.6 em]
\item[\it Case~D:] In the equation for the TOV potential,
  Eq.~(\ref{eq:tov_potential}), $ m_\mathrm{TOV} $ is replaced by
  $ \frac{1}{2} (m_\mathrm{TOV} + m_\mathrm{g}) $. Here a Newtonian
  gravitational mass is defined as
  $ m_\mathrm{g} = m_\mathrm{r} - m_\mathrm{b} $ with the rest mass
  $ m_\mathrm{r} = 4 \pi \int_0^r \mathrm{d}r' \, r'^2 \rho $ and the
  mass equivalent of the binding energy
  $ m_\mathrm{b} = 2 \pi |\int_0^r \mathrm{d}r' \, r'^2 \rho \, \Phi| $.
  As $ m_\mathrm{g} < m_\mathrm{TOV} $, the strength of the potential
  is reduced.
  \\ [-0.6 em]
\item[\it Case~E:] Both in the equation for the TOV potential,
  Eq.~(\ref{eq:tov_potential}), and the equation for the TOV mass,
  Eq.~(\ref{eq:tov_mass}), we set $ e = 0 $.
  \\ [-0.6 em]
\item[\it Case~F:] In the equation for the TOV potential,
  Eq.~(\ref{eq:tov_potential}), we set $ \Gamma = 1 $. As
  $ \Gamma < 1 $ otherwise, this weakens the potential.
  \\ [-0.6 em]
\item[\it Case~G:] In the expression for $ \Gamma $,
  Eq.~(\ref{eq:gamma_factor}), the velocity is set to zero, $ v = 0 $.
  Hence, $ \Gamma^{-2} $ increases in Eq.~(\ref{eq:tov_potential}).
  This modification is used to also test a potential which is even
  stronger than the unmodified TOV potential.
\end{description}
In addition to these cases with a modified version of the TOV
potential, we use the following notations:

\begin{description}
\item[\it Case~N:] This denotes the purely Newtonian runs with
  ``regular'' Newtonian potential.
  \\ [-0.6 em]
\item[\it Case~R:] This is the ``reference'' case with the TOV
  potential as defined by Eq.~(\ref{eq:tov_potential}).
  \\ [-0.6 em]
\item[\it Case~GR:] This case refers to fully relativistic simulations
  with either the code \textsc{CoCoNuT} or the
  \textsc{Agile-BoltzTran} neutrino radiation-hydrodynamics code of
  the Oak Ridge-Basel collaboration.
\end{description}

Note that setting the internal energy density $e$ to zero in Case~B is
unambiguous when a simple EoS is used and the particle rest masses are
conserved. In general, however, particles can be created and
destroyed, or bound states can be formed (e.g., in pair annihilation
processes or nuclear reactions, respectively). Then only the sum of
the rest mass energy and internal energy per nucleon -- both appear in
Eqs.~(\ref{eq:tov_potential}, \ref{eq:tov_mass}) only combined in form
of the ``relativistic energy'' per unit of mass, $ (\rho + e) / \rho $
-- is well defined, but not the individual parts. Therefore there
exists ambiguity with respect to which  contribution to the energy is
set to zero. In order to assess a possible sensitivity of the core
collapse results to this ambiguity, we tried
two different variants of Case~B in our \textsc{Vertex} simulations
with microphysical EoS. On the one hand we used $ \widetilde{e} = E =
F = 0 $ in Eq.~(\ref{eq:tov_mass}), with $ \widetilde{e} $ being the
internal energy density plus an energy normalization given by the EoS
of \citet{lattimer_91_a},
$ \widetilde{e} = (\rho + e) - \rho \, (m_\mathrm{n} - \Delta) / m_\mathrm{u} $
(where $ m_\mathrm{u} = 1.66 \times 10^{-24} \mathrm{\ g} $ is the
atomic mass unit, $ m_\mathrm{n} $ the neutron rest mass, and
$ \Delta = 8.8 \mathrm{ MeV} $). On the other hand we tested
$ e^\ast = E = F = 0 $ with $ e^\ast $ being $ \widetilde{e} $ without
this energy normalization, i.e.\
$ e^\ast = (\rho + e) - \rho \, m_\mathrm{n} / m_\mathrm{u} $.
Nucleons are then assumed to contribute to the TOV mass,
Eq.~(\ref{eq:tov_mass}), with the vacuum rest mass of the neutron,
increasing the mass integral and reducing $ \Gamma $
(Eq.~(\ref{eq:gamma_factor}) relative to the first case, thus making
the effective relativistic potential a bit stronger again. In order to
compensate for this we also set $ p_\nu = 0 $ in the TOV potential,
Eq.~(\ref{eq:tov_potential}). Both variants are found to yield
extremely similar results and we therefore will discuss only one of
them (the first variant) as Case~B for the \textsc{Vertex} simulations.

Ideally, a Newtonian simulation with an effective relativistic
potential not only yields a solution of the TOV structure equations
for an equilibrium state (as does Case~R), but in addition closely
reproduces the results from a relativistic simulation (Case~GR) during
a dynamic evolution. Applying the modifications of the TOV potential
listed above, we find that Cases~A to~D yield improved results as
compared to Case~R, while Cases~E to~G either weaken the potential
too much or are very close to Case~R. These findings are detailed in
Sect.~\ref{sec:spherical_simulations}
and~\ref{sec:rotating_simulations}.

% END SUBSECTION VARIANTS OF EFFECTIVE POTENTIAL
%%%%%%%%%%%%%%%%%%%%%%%%%%%%%%%%%%%%%%%%%%%%%%%%%%%%%%%%%%%%%%%%%
%%%%%%%%%%%%%%%%%%%%%%%%%%%%%%%%%%%%%%%%%%%%%%%%%%%%%%%%%%%%%%%%%

%%%%%%%%%%%%%%%%%%%%%%%%%%%%%%%%%%%%%%%%%%%%%%%%%%%%%%%%%%%%%%%%%
%%%%%%%%%%%%%%%%%%%%%%%%%%%%%%%%%%%%%%%%%%%%%%%%%%%%%%%%%%%%%%%%%
% SUBSECTION THEORETICAL MOTIVATION

\subsection{Theoretical motivation}
\label{subsec:theoretical_motivation}

There are (at least) two basic requirements which appear
desirable for an effective relativistic potential in a
Newtonian simulation. Firstly, the far field limit
of the fully relativistic treatment should be approximated
reasonably well in order to follow the long-term accretion
of the neutron star and the associated growth of its baryonic
mass. Secondly, the hydrostatic structure of the neutron star
should well fit the solution of the TOV equations.

The second point will be discussed in detail in
Sect.~\ref{subsec:simple_core_collapse}.
A closer consideration of the first point suggests the
modified effective relativistic potential of Case~A as promising, and
in fact it turns out to be the most preferable choice concerning
consistency and quality of the results. The other cases listed in
Sect.~\ref{subsec:improved_potentials} are mostly ad hoc
modifications of the original effective relativistic potential of
Eqs.~(\ref{eq:tov_potential})--(\ref{eq:tov_mass}) (Case~R) with
the aim to reduce its strength, which was found to overestimate
the effects of gravity compared to fully relativistic simulations
in previous work (Liebend\"orfer et al.~2005). These cases are
also discussed here for reasons of comparison and completeness.

In Eq.~(\ref{eq:tov_mass}) the hydrodynamic quantities (like rest-mass
density $ \rho $ plus extra terms) are integrated over volume. In the
Newtonian treatment there is no distinction between coordinate volume
and local proper volume. Performing the  integral of
Eq.~(\ref{eq:tov_mass}) therefore leads to a mass -- used as the
mass which produces the gravitational potential in
Eqs.~(\ref{eq:tov_potential}) and (\ref{eq:effective_potential}) --
which is {\em larger} than the baryonic mass,
$ m_\mathrm{b} = 4 \pi \int \mathrm{d}r' \, r'^2 \rho $. In
particular, it is also larger than the gravitational mass in a
consistent relativistic treatment, which is the volume integral of the
total energy density  and includes the negative gravitational
potential energy of the  compact object. The latter reduces the
gravitational mass relative  to the baryonic mass by the gravitational
binding energy of the star \citep[see, e.g.,][page~125 for a
corresponding discussion]{shapiro_83_a}. Therefore, the effective
relativistic potential introduced by \citet{rampp_02_a} [our Case~R,
Eqs.~(\ref{eq:tov_potential}--\ref{eq:tov_mass})] cannot properly
reproduce the far field limit of the relativistic case and thus
overestimates the effects of gravity. This particularly applies to the
infall velocities of the stellar gas ahead of the supernova shock, as
shown in \citet{liebendoerfer_05_a}.

Introducing an extra factor $ \Gamma $ in the integral of
Eq.~(\ref{eq:tov_mass}) for the TOV mass is motivated by the following
considerations (where for reasons of simplicity contributions from
neutrinos, though important, are neglected and spherical symmetry is
assumed): In the relativistic treatment the total (gravitating) mass
of the star is given as
\begin{equation}
  m_\mathrm{g} = 4 \pi \int_0^\infty \!\! \mathrm{d}r' \,
  r'^2 (\rho + e),
  \label{eq:gravitating_mass}
\end{equation}
where $ \mathrm{d}V' = 4 \pi \mathrm{d}r' \, r'^2 $ is the coordinate
volume element, whereas the baryonic mass is
\begin{equation}
  m_\mathrm{b} = 4 \pi \int_0^\infty \!\! \mathrm{d}r' \,
  r'^2 \Gamma^{-1} \rho,
  \label{eq:baryonic_mass}
\end{equation}
with $ \mathrm{d}{\cal V}' = 4 \pi \mathrm{d}r' r'^2 \Gamma^{-1} $
being the local proper volume element. The factor $ \Gamma^{-1} > 1 $
in the integrand of $ m_\mathrm{b} $ thus ensures that
$ m_\mathrm{b} > m_\mathrm{g} $. The integral for $ m_\mathrm{g} $ can
also be written as
\begin{equation}
  m_\mathrm{g} = 4 \pi \int_0^\infty \!\! \mathrm{d}r' \,
  r'^2 \, \frac{\Gamma}{\Gamma} \, (\rho + e) =
  \int_0^\infty \!\! \mathrm{d}{\cal V}' \, \Gamma \, (\rho + e).
  \label{eq:gravitating_mass_alt}
\end{equation}
Since in Newtonian hydrodynamics no distinction is made between
coordinate and proper volumes, one may identify
$ \mathrm{d}{\cal V}' \equiv \mathrm{d}V' $, consistent with the rest
of our Newtonian code. This leaves the additional factor $ \Gamma $ in
the integrand of Eq.~(\ref{eq:gravitating_mass_alt}), leading to a
redefined TOV mass used for computing the effective relativistic
potential in Case~A,
\begin{equation}
  \widetilde{m}_\mathrm{TOV}(r) = 4 \pi \int_0^r \mathrm{d}r' \,
  r'^2 \Gamma \left( \rho + e + E + \frac{v F}{\Gamma} \right).
  \label{eq:tov_mass_redefined}
\end{equation}
The fact that a factor $ \Gamma^{-1} $ in the volume integral
establishes the relation between gravitating mass,
Eq.~(\ref{eq:gravitating_mass}), and baryonic mass,
Eq.~(\ref{eq:baryonic_mass}), in the relativistic case suggests that
the factor $\Gamma < 1$ in Eq.~(\ref{eq:tov_mass_redefined}) might
lead to a suitable reduction of the overestimated effective potential
that results when the original TOV mass of Eq.~(\ref{eq:tov_mass}) is
used in Eqs.~(\ref{eq:tov_potential}) and
(\ref{eq:gamma_factor}). Indeed, a comparison of the integral of
Eq.~(\ref{eq:tov_mass_redefined}) for large $r$ with the rest-mass
energy of a neutron star reduced by its binding energy at time $t$
(computed from the emitted neutrino energy,
$ \int_0^t \mathrm{d}t' \, L_\nu(t') $ with $ L_\nu $ being the
neutrino luminosity) reveals very good agreement.

The arguments given above only provide a heuristic justification
for the manipulation of the TOV potential proposed in Case~A. A deeper
theoretical understanding and more rigorous analytical analysis of its
consequences and implications is certainly desirable, but beyond the
scope of the present paper. We plan to return to this question in
future work.

% END SUBSECTION THEORETICAL MOTIVATION
%%%%%%%%%%%%%%%%%%%%%%%%%%%%%%%%%%%%%%%%%%%%%%%%%%%%%%%%%%%%%%%%%
%%%%%%%%%%%%%%%%%%%%%%%%%%%%%%%%%%%%%%%%%%%%%%%%%%%%%%%%%%%%%%%%%

% END SECTION GRAVITATIONAL POTENTIAL
%%%%%%%%%%%%%%%%%%%%%%%%%%%%%%%%%%%%%%%%%%%%%%%%%%%%%%%%%%%%%%%%%
%%%%%%%%%%%%%%%%%%%%%%%%%%%%%%%%%%%%%%%%%%%%%%%%%%%%%%%%%%%%%%%%%

%%%%%%%%%%%%%%%%%%%%%%%%%%%%%%%%%%%%%%%%%%%%%%%%%%%%%%%%%%%%%%%%%
%%%%%%%%%%%%%%%%%%%%%%%%%%%%%%%%%%%%%%%%%%%%%%%%%%%%%%%%%%%%%%%%%
% SECTION NUMERICAL METHODS

\section{Numerical methods}
\label{sec:numerical_methods}

In the following we briefly introduce the two numerical codes used for
the simulations presented in this work. The codes are based on
state-of-the-art numerical methods for hydrodynamics and neutrino
transport coupled to gravity, and were used to simulate stellar
core collapse and supernova explosions.

%%%%%%%%%%%%%%%%%%%%%%%%%%%%%%%%%%%%%%%%%%%%%%%%%%%%%%%%%%%%%%%%%
%%%%%%%%%%%%%%%%%%%%%%%%%%%%%%%%%%%%%%%%%%%%%%%%%%%%%%%%%%%%%%%%%
% SUBSECTION PURELY HYDRODYNAMICS

\subsection{Hydrodynamic code without transport}
\label{subsec:hydro_code}

The purely hydrodynamic simulations are performed with the code
\textsc{CoCoNuT} developed by \citet{dimmelmeier_02_a,
dimmelmeier_02_b} with a metric solver based on spectral methods as
described in \citet{dimmelmeier_05_a}. The code optionally uses either
a Newtonian gravitational potential (or an effective relativistic
potential, see Sect.~\ref{sec:effective_potential}), or the general
relativistic field equations for a curved spacetime in the ADM
3\,+\,1-split under the assumption of the conformal flatness condition
(CFC) for the three-metric. The (Newtonian or general relativistic)
hydrodynamic equations are consistently formulated in conservation
form, and are solved by high-resolution shock-capturing schemes based
upon state-of-the-art Riemann solvers and third-order
cell-reconstruction procedures. Neutrino transport is not included in
the code. A simple hybrid ideal gas EoS is used that consists of a
polytropic contribution describing the degenerate electron pressure
and (at supranuclear densities) the pressure due to repulsive nuclear
forces, and a thermal contribution which accounts for the heating of
the matter by shocks:
\begin{equation}
  P = P_\mathrm{p} + P_\mathrm{th},
  \label{eq:hybrid_eos}
\end{equation}
where
\begin{equation}
  P_\mathrm{p} = K \rho^{\gamma},
  \qquad
  P_\mathrm{th} = \rho \epsilon_\mathrm{th} (\gamma_\mathrm{th} - 1),
  \label{eq:hybrid_eos_terms}
\end{equation}
and $ \epsilon_\mathrm{th} = \epsilon - \epsilon_\mathrm{p} $. The
polytropic specific internal energy $ \epsilon_\mathrm{p} $ is
determined from $ P_\mathrm{p} $ by the ideal gas relation in
combination with continuity conditions in the case of a discontinuous
$ \gamma $. In that case, the polytropic constant $ K $ also has to be
adjusted \citep[for more details, see][]{dimmelmeier_02_a,
  janka_93_a}.

The \textsc{CoCoNuT} code utilizes Eulerian spherical coordinates
$ \{r, \theta, \varphi \} $, and thus axially or spherically symmetric
configurations can be easily simulated. For the core collapse
simulations discussed in Sects.~\ref{subsec:simple_core_collapse}
and~\ref{sec:rotating_simulations}, the finite difference grid consists
of 200 logarithmically spaced radial grid points with a central
resolution of $ 500 \mathrm{\ m} $. A small part of the
grid covers an artificial low-density atmosphere extending beyond the
core's outer boundary. The spectral grid of the metric solver is split
into 6 radial domains with 33 collocation points each. In order
to be able to accurately follow the dynamics, the domain boundaries
adaptively contract towards the centre during the collapse, as
described in \citet{dimmelmeier_05_a}. For the migration
test (Sect.~\ref{subsec:migration_test}) the finite difference grid
consists of 250 logarithmically spaced radial grid points, of which 60
and 190 cover the neutron star and the atmosphere, respectively. For
the two-dimensional simulations of rotational core collapse
(Sect.~\ref{sec:rotating_simulations}) the finite difference and
spectral grids are extended by 30 and 17 equidistant angular grid
points, respectively.

Even when using spectral methods the calculation of the spacetime
metric is computationally expensive. Hence, in the relativistic
spherical (rotational) core collapse simulations the metric is updated
only once every 10/1/50 (100/10/50) hydrodynamic time steps
before/during/after core bounce, and extrapolated in between. In the
migration test the update is performed every 10th step during the
entire evolution. The numerical adequacy of this procedure is tested
and discussed in detail in \citet{dimmelmeier_02_a}.

The quality of the CFC approximation has been tested comprehensively
in the context of supernova core collapse and for neutron star models
\citep{cook_96_a, dimmelmeier_02_a, shibata_04_a, dimmelmeier_05_a,
cerda_04_a}. For single stellar configurations the CFC approximation
is very accurate as long as neither the compactness of the star is
extremely relativistic nor its rotation is too fast and
differential. Both criteria are well fulfilled during supernova core
collapse. Concerning core collapse simulations of rapidly and strongly
differentially rotating configurations at very high densities, we
recently found that the CFC approximation yields excellent agreement
with formulations solving the exact spacetime metric even in this
extreme regime \citep{ott_05_a}.

As the three-metric of any spherically symmetric spacetime can be
written in a conformally flat way, the spherical CFC simulations
presented in Sect.~\ref{subsec:simple_core_collapse}
and~\ref{subsec:migration_test} involve no approximation, i.e., the
exact Einstein equations are solved in these cases.

% END SUBSECTION PURELY HYDRODYNAMICS
%%%%%%%%%%%%%%%%%%%%%%%%%%%%%%%%%%%%%%%%%%%%%%%%%%%%%%%%%%%%%%%%%
%%%%%%%%%%%%%%%%%%%%%%%%%%%%%%%%%%%%%%%%%%%%%%%%%%%%%%%%%%%%%%%%%

%%%%%%%%%%%%%%%%%%%%%%%%%%%%%%%%%%%%%%%%%%%%%%%%%%%%%%%%%%%%%%%%%
%%%%%%%%%%%%%%%%%%%%%%%%%%%%%%%%%%%%%%%%%%%%%%%%%%%%%%%%%%%%%%%%%
% SUBSECTION NEUTRINO TRANSPORT

\subsection{Hydrodynamic code with neutrino transport}
\label{subsec:transport_code}

The \textsc{Vertex} neutrino-hydrodynamics code solves the
conservation equations of Newtonian hydrodynamics in their Eulerian
and conservative form using the PPM-based \textsc{Prometheus} code
\citep{fryxell_90_a}. The neutrino transport is treated by determining
iteratively a solution of the coupled set of moments equations and
Boltzmann equation, achieving closure by a variable Eddington factor
method \citep{rampp_02_a}. See the latter reference also for details
about the coupling between the hydrodynamics and neutrino transport
parts of the \textsc{Vertex} code. The models presented in this paper
are computed with the equation of state of \citet{lattimer_91_a}, in
agreement with the choice of input physics in the work of
\citet{liebendoerfer_05_a}, where results obtained with the
\textsc{Vertex} code were compared with those from the Newtonian and
fully relativistic calculations with the \textsc{Agile-BoltzTran} code
of the Oak Ridge-Basel collaboration.

In order to compare the results presented here with those of the
calculations of \citet{liebendoerfer_05_a} we used the same set of
neutrino interaction rates as picked for Model G15 in
\citet{liebendoerfer_05_a}, and exactly the same parameters for the
numerical setup (e.g., the grids for hydrodynamics and neutrino
transport). Information about this setup can be found in
\citet{liebendoerfer_05_a}. The initial model for our calculations is
the $ 15 M_\odot $ progenitor model ``s15s7b2'' from
\citet{woosley_95_a}.

Since solving the neutrino transport problem is computationally quite
expensive we performed calculations only for Cases~A, B, and~F
(as defined in Sect.~\ref{subsec:improved_potentials}) with the
\textsc{Vertex} code. The quality of these results is then compared to
the fully relativistic treatment of the \textsc{Agile-BoltzTran} code.

% END SUBSECTION NEUTRINO TRANSPORT
%%%%%%%%%%%%%%%%%%%%%%%%%%%%%%%%%%%%%%%%%%%%%%%%%%%%%%%%%%%%%%%%%
%%%%%%%%%%%%%%%%%%%%%%%%%%%%%%%%%%%%%%%%%%%%%%%%%%%%%%%%%%%%%%%%%

%%%%%%%%%%%%%%%%%%%%%%%%%%%%%%%%%%%%%%%%%%%%%%%%%%%%%%%%%%%%%%%%%
%%%%%%%%%%%%%%%%%%%%%%%%%%%%%%%%%%%%%%%%%%%%%%%%%%%%%%%%%%%%%%%%%
% SUBSECTION HYDRO EQUATIONS AND POTENTIAL IMPLEMENTATION

\subsection{Hydrodynamics and implementation of the effective
  relativistic potential}
\label{subsec:code_implementation}

The implementation of an effective relativistic gravitational
potential or of an effective relativistic gravitational force as its
derivative into existing Newtonian hydrodynamics codes is
straightforward and does not differ from the use of the Newtonian
potential or force.

The equations solved by the two codes used for our simulations are
described in much detail in previous publications \citep[see,
e.g.,][and references therein]{rampp_02_a, mueller_95_a,
  dimmelmeier_02_a, dimmelmeier_05_a}. The implementation of the
source terms of the gravitational potential as discussed in
\citet{mueller_95_a} is also applied for the handling of the
effective relativistic potentials investigated in this work. The only
specific feature here is the mutual dependence of $ \Gamma $ and
$ m_\mathrm{TOV} $, which is either accounted for by a rapidly
converging iteration or by taking $ \Gamma $ from the old time step in
the update of $ m_\mathrm{TOV} $, when the changes during the time
steps are sufficiently small.

Since the actual form of the gravitational source term is unspecified
in the conservation laws of fluid dynamics, the Newtonian potential
can be replaced by the effective relativistic potentials investigated
in the current work in a technically straightforward way. Solving an
equation for the internal plus kinetic energy (as in our codes)
requires a treatment of the gravity source term in this equation that
is consistent with its implementation in the equation of momentum.

Of course, the effective potential must be investigated concerning its
consequences for the conservation of momentum and energy. Since a
potential constructed according to Eqs.~(\ref{eq:tov_potential},
\ref{eq:tov_mass}) does not satisfy the Poisson equation, the momentum
equation cannot be cast into a conservation form \citep[cf.][Part~I,
Chapter~4]{shu_92_a}. As a consequence, the total linear momentum is
strictly conserved only when certain assumptions about the symmetry of
the matter distribution are made, for example in the case of spherical
symmetry or axially symmetric configurations with equatorial symmetry,
or when only one octant is modeled in the three-dimensional
case. In axisymmetric simulations the conservation of specific angular
momentum is fulfilled as well, when using the effective relativistic
potential. In general, however, a sufficient quality of momentum (and
angular momentum) conservation has to be verified by inspecting the
numerical results.

The long-range nature of gravity prohibits to have an equation in pure
conservation form for the total energy, i.e., for the sum of internal,
kinetic, and gravitational energy \citep[][Part~I,
Chapter~4]{shu_92_a}. In contrast to the Newtonian case, however, our
effective relativistic potential does also not allow one to derive a
conservation equation for the total energy integrated over all
space. Monitoring global energy conservation in a simulation with
effective relativistic potential therefore requires integration of the
gravitational source terms over all cells in time. If the local
effects of relativistic gravity are convincingly approximated -- as
measured by good agreement with static solutions of the TOV equation
and with fully relativistic, dynamical simulations -- there is
confidence that the integrated action of the employed gravitational
source term approximates well also the global conversion between total
kinetic, internal, and gravitational energies found in a relativistic
simulation.

The recipes for approximating general relativity should be applicable
equally well in hydrodynamic codes different from our (Eulerian) PPM
schemes, provided the effects of gravity are consistently treated in
the momentum and energy equations. The proposed effective potentials
are intended to yield a good representation of the effects of
relativistic gravity in particular in the context of stellar core
collapse and neutron star formation. For our approximation to work
well, the fluid flow should be subrelativistic. The numerical tests
described in the following sections show that velocities up to about
20\% of the speed of light are unproblematic.

% END SUBSECTION HYDRO EQUATIONS AND POTENTIAL IMPLEMENTATION
%%%%%%%%%%%%%%%%%%%%%%%%%%%%%%%%%%%%%%%%%%%%%%%%%%%%%%%%%%%%%%%%%
%%%%%%%%%%%%%%%%%%%%%%%%%%%%%%%%%%%%%%%%%%%%%%%%%%%%%%%%%%%%%%%%%

%%%%%%%%%%%%%%%%%%%%%%%%%%%%%%%%%%%%%%%%%%%%%%%%%%%%%%%%%%%%%%%%%
%%%%%%%%%%%%%%%%%%%%%%%%%%%%%%%%%%%%%%%%%%%%%%%%%%%%%%%%%%%%%%%%%
% SECTION SIMULATIONS RESULTS 1D

\section{Simulations in spherical symmetry}
\label{sec:spherical_simulations}

%%%%%%%%%%%%%%%%%%%%%%%%%%%%%%%%%%%%%%%%%%%%%%%%%%%%%%%%%%%%%%%%%
%%%%%%%%%%%%%%%%%%%%%%%%%%%%%%%%%%%%%%%%%%%%%%%%%%%%%%%%%%%%%%%%%
% SUBSECTION SIMULATIONS RESULTS 1D HYDRO

\subsection{Supernova core collapse with simplified equation of state}
\label{subsec:simple_core_collapse}

We first test the quality of the TOV potential and its various
modifications as an effective relativistic potential in the moderately
relativistic regime by evolving models of supernova core collapse with
the simple EoS described in Sect.~\ref{subsec:hydro_code}. Using the
\textsc{CoCoNuT} code we are able to perform a direct comparison with
a Newtonian version of the code (employing either the Newtonian, TOV,
or modified TOV potential) against the relativistic version in a
curved spacetime.

\begin{figure}
  \resizebox{\hsize}{!}
  {\includegraphics{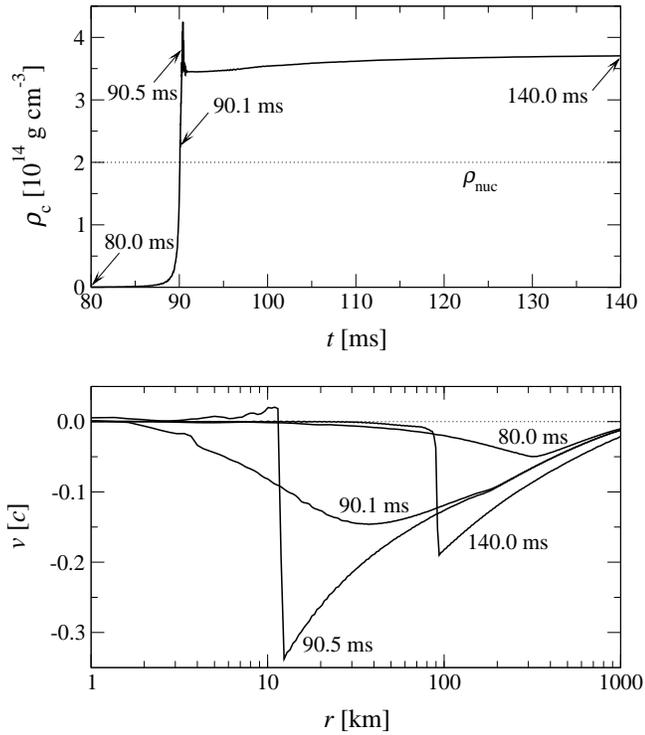}}
  \caption{Time evolution of the central density $ \rho_\mathrm{c} $
    (top panel) and radial profiles of the velocity $ v $ (bottom
    panel) for the accretion shock model AS in relativistic gravity
    (Case~GR). The velocity profiles are snapshots long before bounce,
    during bounce shortly before and after maximum density, and long
    after bounce, respectively. At $ t = 140.0 \mathrm{\ ms} $ the
    prompt shock has turned into an accretion shock.}
  \label{fig:accretion_shock_collapse}
\end{figure}

The initial models are polytropes, which mimic an iron core supported
by electron degeneracy pressure, with a central density
$ \rho_{\rm c\,i} = 10^{10} {\rm\ g\ cm}^{-3} $ and EoS parameters
$ \gamma_\mathrm{i} = 4 / 3 $ and $ K = 4.897 \times 10^{14} $ (in cgs
units). To initiate the collapse, the initial adiabatic index is
reduced to $ \gamma_1 < \gamma_\mathrm{i} $. At densities
$ \rho > \rho_{\rm nuc} \equiv 2.0 \times 10^{14} {\rm\ g\ cm}^{-3} $
the adiabatic index is increased to $ \gamma_2 \gsim 2.5 $ to
approximate the stiffening of the EoS. This leads to a rebound of the
core (``core bounce''). The inner part of the core comes to a halt,
while a prompt hydrodynamic shock starts to propagate outward. Since
the matter model of the \textsc{CoCoNuT} code does not account for
neutrino cooling, the compact collapse remnant (corresponding to the
hot proto-neutron star) cannot cool and shrink, but for some models
only slightly grows in mass due to accretion. These stages of the
evolution can be seen in the top panel of
Fig.~\ref{fig:accretion_shock_collapse}, where the time evolution of
the central density $ \rho_\mathrm{c} $ is plotted for a typical core
collapse model.

As it is known from numerical models with sophisticated microphysics
that the prompt hydrodynamic shock turns into an accretion shock
shortly after core bounce, we choose the EoS parameters
$ \gamma_1 = 1.325 $, $ \gamma_2 = 3.0 $, and
$ \gamma_\mathrm{th} = 1.2 $ in our accretion shock model AS
(Fig.~\ref{fig:accretion_shock_collapse}) to reproduce such a
behaviour. The conversion of the prompt shock into an accretion shock
at late times is reflected by the radial velocity profiles plotted in
the bottom panel of Fig.~\ref{fig:accretion_shock_collapse}.

\begin{figure}
  \resizebox{\hsize}{!}
  {\includegraphics{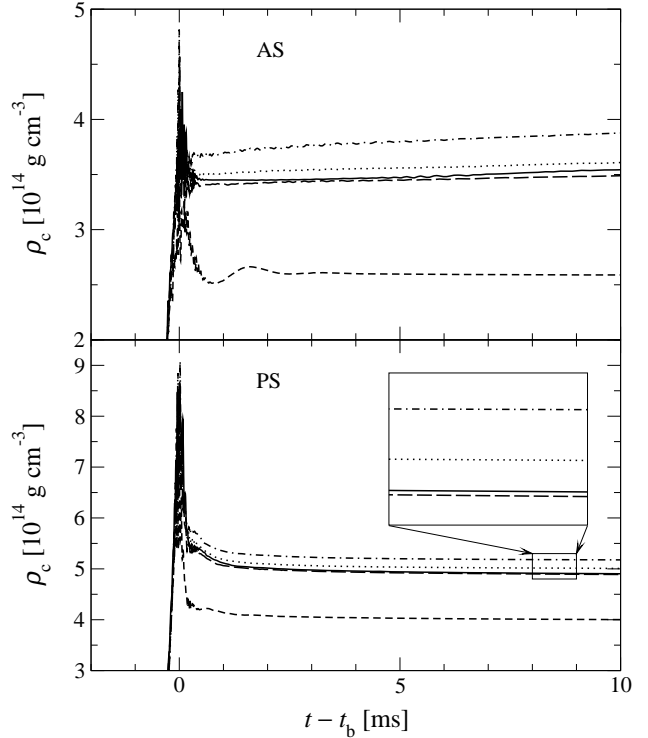}}
  \caption{Time evolution of the central density $ \rho_\mathrm{c} $
    for the accretion shock model AS (top panel) and prompt shock
    model PS (bottom panel). The regular Newtonian simulation (Case~N;
    dashed line) yields a smaller $ \rho_\mathrm{c} $ compared to the
    relativistic one (Case~GR; solid line). If a TOV potential is used
    (Case~R; dashed-dotted line), $ \rho_\mathrm{c} $ is too high,
    while modifications~A (dotted line) and~B (long-dashed line) of
    the TOV potential give results close to the relativistic one.}
  \label{fig:comparison_density_evolution}
\end{figure}

When recalculating Model AS with regular Newtonian gravity (Case~N),
the central density remains below the one for relativistic gravity
(Case~GR) during the entire evolution (see top panel of
Fig.~\ref{fig:comparison_density_evolution}). This behaviour was
already observed in the rotational core collapse simulations of
\citet{dimmelmeier_02_a}. Replacing the regular Newtonian potential
(Case~N) by the TOV potential (Case~R) yields the opposite effect,
i.e., the central density is too high throughout the
evolution. However, when using modifications~A or~B of the TOV
potential, the behaviour of $ \rho_\mathrm{c} $ agrees very well with
that observed in the relativistic simulation (Case~GR), indicating that
the dynamics of the core collapse with this effective relativistic
potential is similar to that of the relativistic case.

When applying the same methods to a model with a prompt shock which
continues to propagate, we observe the same qualitative behaviour (see
bottom panel of Fig.~\ref{fig:comparison_density_evolution}). For this
Model PS the EoS parameters are set to $ \gamma_1 = 1.300 $,
$ \gamma_2 = 2.0 $, and $ \gamma_\mathrm{th} = 1.5 $, respectively.
Both during core bounce and at late times (see magnifying inset) the
Newtonian simulations with modifications~A and~B of the TOV potential
also yield results which agree noticeably better with the relativistic
one (Case~GR) than with the TOV potential (Case~R), and obviously also
much better than with the regular potential (Case~N).

The time plotted in Fig.~\ref{fig:comparison_density_evolution} is
$ t - t_\mathrm{b} $, where $ t_\mathrm{b} $ is the time of bounce for
each simulation. We point out that the time coordinate $ t $ in the
relativistic spacetime is the time of an observer being at rest at
infinity, and thus we can directly relate it to the time used in the
Newtonian simulation (with regular or effective relativistic
potential). Moreover, at core bounce the differences between the
relativistic coordinate time $ t $ and the relativistic proper time
$ t_\mathrm{p\,c} = \int \alpha_\mathrm{c} \, \mathrm{d}t $ of the
fluid element at the center, where $ \alpha $ is the ADM lapse
function, are negligible for the gravitational fields encountered in
our core collapse models. In addition, shortly after core bounce the
compact collapse remnant reaches an approximate hydrostatic
equilibrium with nearly constant $ \rho_\mathrm{c} $. Therefore, for
the models considered here the time evolution of the central density
in proper time $ t_\mathrm{p\,c} $ would look similar to the one in
coordinate time $ t $.

The discussion up to now has only been concerned with the local
quantity $ \rho_\mathrm{c} $. While this is a good measure for the
global collapse dynamics, a direct comparison of, e.g., radial
profiles at certain evolution times is more powerful.

\begin{figure}
  \resizebox{\hsize}{!}
  {\includegraphics{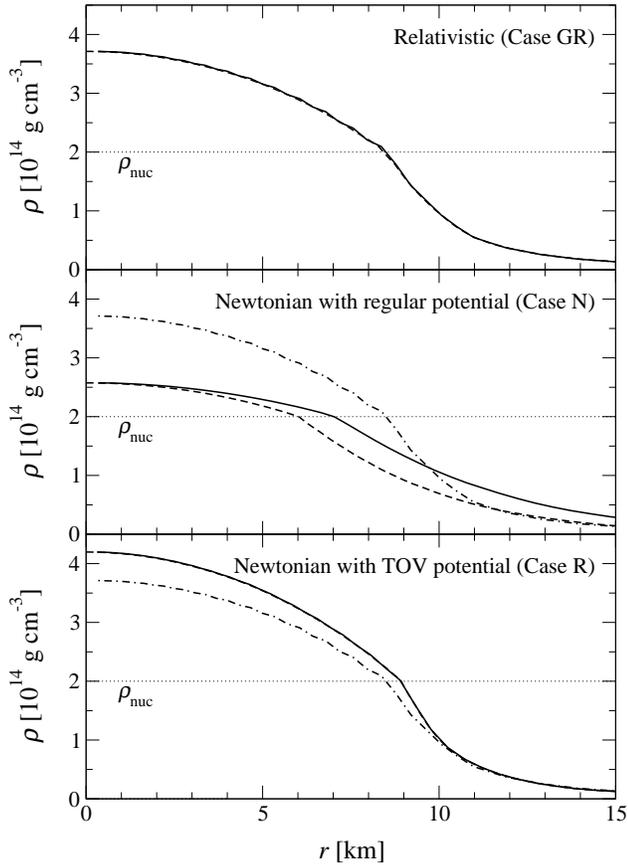}}
  \caption{Comparison of radial profiles of the density $ \rho $ from
    the hydrodynamic evolution of Model AS at late times (solid line)
    and from a solution of the TOV structure equations with identical
    central density (dashed line). In both the relativistic simulation
    (Case~GR; top panel) and the Newtonian simulation with TOV
    potential (Case~R; bottom panel) the two curves are almost
    indistinguishable, while the regular Newtonian simulation (Case~N;
    centre panel) yields a large mismatch. For comparison, the
    relativistic profile (solid line in the top panel) is also plotted
    in dashed-dotted line style in the lower two panels.}
  \label{fig:tov_comparison_density_profiles_1}
\end{figure}

In Fig.~\ref{fig:tov_comparison_density_profiles_1} radial profiles of
the density $ \rho $ for the core collapse model AS are compared at
``late'' times ($ t = 150 \mathrm{\ ms} $, corresponding to about
$ 60 \mathrm{\ ms} $ after core bounce). As this model exhibits
  no post-bounce mass accretion, shortly after core bounce the compact
remnant has acquired a new equilibrium state with essentially
vanishing radial velocity. Its density profile can thus be very well
described by that of a TOV solution constructed with equal central
density $ \rho_\mathrm{c} $ and an ``effective'' one-parameter EoS,
$ P = P (\rho) $, extracted from the two-parameter EoS,
$ P = P (\rho, \epsilon) $, defined by Eqs.~(\ref{eq:hybrid_eos},
\ref{eq:hybrid_eos_terms}). Details of this procedure are given in
Appendix~\ref{app:tov_solution}. The top panel of
Fig.~\ref{fig:tov_comparison_density_profiles_1} shows that this
behaviour is indeed confirmed by our simulations. The density profile
obtained from a dynamic simulation with relativistic gravity (Case~GR)
and the one obtained from solving the TOV structure equations agree
extremely well both at supranuclear and subnuclear densities. Only
ahead of the outward propagating shock (off scale in
Fig.~\ref{fig:tov_comparison_density_profiles_1}) where matter is
still falling inward, i.e., where the radial velocity is not
negligible, a disagreement can be observed.

When using regular Newtonian gravity (Case~N; see centre panel of
Fig.~\ref{fig:tov_comparison_density_profiles_1}) the situation is
different in two aspects. First, due to the shallower Newtonian
gravitational potential, the density is significantly smaller in the
central parts of the compact remnant compared to relativistic gravity
(Case~GR). However, beyond $ r \sim 10 \mathrm{\ km} $ the density in
the Newtonian model exceeds that of the relativistic simulation, which
is the density crossing effect observed and discussed in detail by
\citet{dimmelmeier_02_b}. The weaker regular Newtonian potential thus
yields a remnant which is less compact and dense in the centre, but
relatively denser in the outer parts.

\begin{figure}
  \resizebox{\hsize}{!}
  {\includegraphics{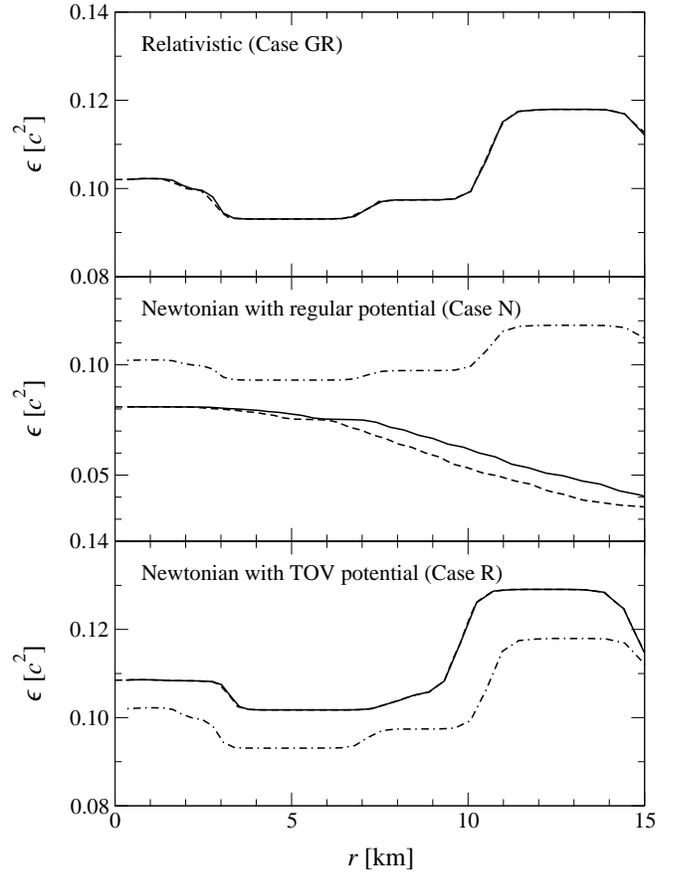}}
  \caption{Same as Fig.~\ref{fig:tov_comparison_density_profiles_1},
    but for radial profiles of the specific internal energy
    $ \epsilon $. The step-like shape of the profile arises from
    contributions of the thermal energy $ \epsilon_\mathrm{th} $ in
    various strengths.}
  \label{fig:tov_comparison_energy_profiles_1}
\end{figure}

Second, when comparing the density profile resulting from the
hydrodynamic evolution with that obtained by solving the TOV structure
equations (for the same $ \rho_\mathrm{c} $), a clear disagreement can
be noticed. As the core reaches a final compactness
$ 2 M_\mathrm{core} / R_\mathrm{core} \sim 0.2 $ as measured by the
ratio of core mass to core radius, a hydrodynamic simulation with
regular Newtonian potential cannot be expected to yield an equilibrium
result which is identical to a solution of the (relativistic) TOV
structure equations.

If on the other hand the Newtonian hydrodynamic simulation is
performed with the TOV potential (Case~R; see bottom panel of
Fig.~\ref{fig:tov_comparison_density_profiles_1}), the final
equilibrium state is (as in Case~GR) a solution of the TOV structure
equations. However, after core bounce the density profile of
Case~R lies well above that obtained in the relativistic simulation
(Case~GR). Thus Case~GR and Case~R both exhibit post-bounce density
stratifications which are consistent with a relativistic equilibrium
state but otherwise differ noticeably from each other.

Radial profiles of the specific internal energy (see
Fig.~\ref{fig:tov_comparison_energy_profiles_1}) exhibit a similar
behaviour as the density profiles presented in
Fig.~\ref{fig:tov_comparison_density_profiles_1}. As expected, both
the relativistic simulation (Case~GR) and the Newtonian simulation
with TOV potential (Case~R) yield final equilibrium profiles which
closely coincide with the corresponding solution of the TOV structure
equations (see top and bottom panel, respectively). Again,
analogous to the density, in Case~R the internal energy in the
compact remnant is significantly higher as compared to Case~GR.
In the Newtonian simulation with regular potential (Case~N; centre
panel) the profile deviates clearly from both the corresponding
solution of the TOV equation with equal central value and the
  profile computed with relativistic gravity (Case~GR), and does not
even show the distinctive steps of that profile.

The above findings imply that the final equilibrium state of a
Newtonian core collapse simulation performed with the TOV potential
(Case~R) is consistent with a TOV solution of identical
central density and specific internal energy. However, if
relativistic effects are important, the density stratification and
energy distribution (which in turn determines the ``effective'' EoS as
described in Appendix~\ref{app:tov_solution}) will differ from that of
a consistently relativistic simulation (Case~GR). Presumably the
combination of the relativistic TOV potential (which depends
nonlinearly on the pressure and the internal energy density) with
Newtonian kinematics in Case~R is responsible for this difference in
the stratification of density and specific internal energy in
the compact remnant compared to Case~GR (as seen in the bottom panels
of Figs.~\ref{fig:tov_comparison_density_profiles_1}
and~\ref{fig:tov_comparison_energy_profiles_1}). Currently we cannot
provide a more detailed analysis for understanding the mechanism by
which these discrepancies are caused. We plan to conduct further
research on this issue.

\begin{figure}
  \resizebox{\hsize}{!}
  {\includegraphics{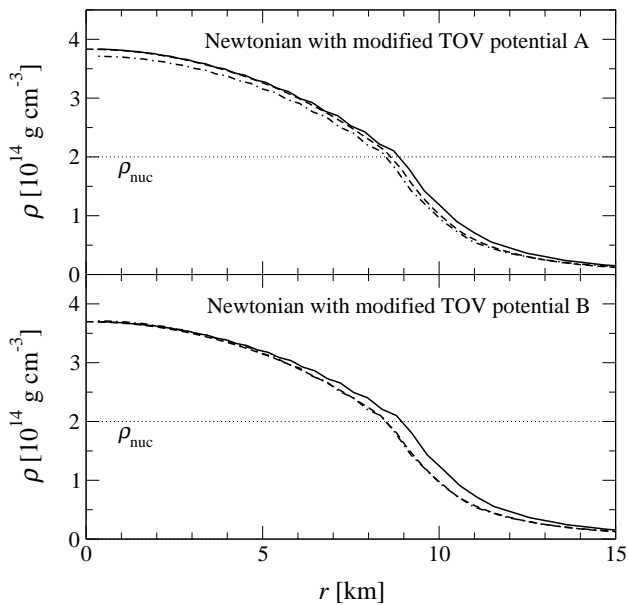}}
  \caption{Comparison of radial profiles of the density $ \rho $ from
    the hydrodynamic evolution of Model AS at late times (solid line)
    and from a solution of the TOV structure equations with identical
    central density (dashed line). In the Newtonian simulations with
    either modification~A (top panel) or~B (bottom panel) of the TOV
    potential, the two curves match well and also agree with the
    result from the relativistic simulation (Case~GR; dashed-dotted
    line).}
  \label{fig:tov_comparison_density_profiles_2}
\end{figure}

\begin{figure}
  \resizebox{\hsize}{!}
  {\includegraphics{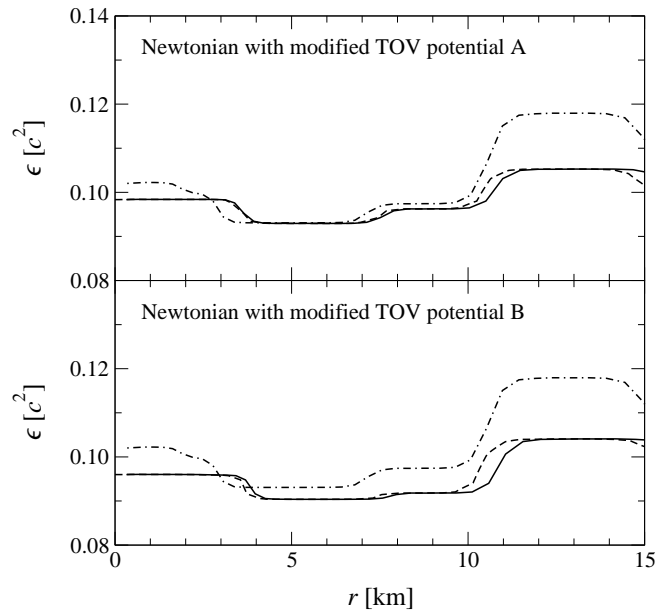}}
  \caption{Same as Fig.~\ref{fig:tov_comparison_density_profiles_2},
    but for radial profiles of the specific internal energy
    $ \epsilon $. The step-like shape of the profile arises from
    contributions of the thermal energy $ \epsilon_\mathrm{th} $ in
    various strengths.}
  \label{fig:tov_comparison_energy_profiles_2}
\end{figure}

As discussed in Sect.~\ref{subsec:improved_potentials}, a modification
of a Newtonian simulation incorporating the effects of relativistic
gravity must not only yield equilibrium states which (approximately)
reproduce a solution of the TOV equations, particularly for compact
matter configurations. More importantly, the density and energy
profiles must also agree as closely as possible with the profiles
obtained from the evolution with a relativistic code (which implies
the previous requirement). Both these requirements are fulfilled for
Newtonian simulations utilizing modification~A and~B of the TOV
potential, as shown for profiles of the density and specific internal
energy in the final equilibrium state of the core collapse model AS
in Figs.~\ref{fig:tov_comparison_density_profiles_2}
and~\ref{fig:tov_comparison_energy_profiles_2}, respectively.
In these cases the profiles agree well with the relativistic result
(Case~GR) not only near the centre (as already observed in the time
evolution of $ \rho_\mathrm{c} $; see top panel of
Fig.~\ref{fig:comparison_density_evolution}) but also throughout the
entire core, with the agreement being much better compared to the TOV
potential (Case~R). Accordingly, the solution of the TOV structure
equations for identical central values yields similar profiles as well.

We conclude that the modifications~A and~B of the TOV potential
combined with Newtonian kinematics provide a very good approximation
of the results obtained with relativistic gravity (Case~GR) in the
strongly dynamic phases during core bounce as well as at post-bounce
times when the compact remnant acquires equilibrium. The same findings
hold for the modifications~C and~D of the TOV potential, although we
do not present results for these cases here. On the other hand,
compared to Cases~A to~D, the modifications~E to~G produce less
accurate results. Thus, we consider them inappropriate for an
effective relativistic potential.

In passing we note that gauge dependences may play an important role
when non-invariant quantities (like, e.g., density profiles) from
simulations with a relativistic code are compared against results from
either a relativistic code using a different formulation of the
spacetime metric or from a Newtonian code. For instance, the radial
coordinate $ r $ used in the profile plots in
Figs.~\ref{fig:tov_comparison_density_profiles_1}
to~\ref{fig:tov_comparison_energy_profiles_2} is the isotropic
coordinate radius of the ADM-CFC metric used in the relativistic
version of the \textsc{CoCoNuT} code \citep[for details,
see][]{dimmelmeier_02_a}. In contrast to this, the TOV
equations~(\ref{eq:tov_equation_pressure}, \ref{eq:tov_equation_mass})
use the standard Schwarzschild-like coordinate radius, which is
equivalent to the circumferential radius. Thus, a coordinate
transformation from the standard Schwarzschild-like radial coordinate
to the isotropic radial coordinate (as specified in
Appendix~\ref{app:coordinate_transformation}) is applied to all
profiles of quantities obtained with the TOV solution or Newtonian
simulations in those figures. Otherwise, the difference in the two
radial coordinates of up to $ \sim 20\% $ would lead to a noticeable
distortion of the radial profiles. We also point out that we neglect
any influence of the difference between Newtonian and relativistic
coordinate time (or proper time) when comparing the profiles at the
same coordinate times, as the compact remnants are in equilibrium at
``late'' times (about $ 60 \mathrm{\ ms} $ after core bounce).

% END SUBSECTION SIMULATIONS RESULTS 1D HYDRO
%%%%%%%%%%%%%%%%%%%%%%%%%%%%%%%%%%%%%%%%%%%%%%%%%%%%%%%%%%%%%%%%%
%%%%%%%%%%%%%%%%%%%%%%%%%%%%%%%%%%%%%%%%%%%%%%%%%%%%%%%%%%%%%%%%%

%%%%%%%%%%%%%%%%%%%%%%%%%%%%%%%%%%%%%%%%%%%%%%%%%%%%%%%%%%%%%%%%%
%%%%%%%%%%%%%%%%%%%%%%%%%%%%%%%%%%%%%%%%%%%%%%%%%%%%%%%%%%%%%%%%%
% SUBSECTION SIMULATIONS RESULTS 1D NEUTRINO TRANSPORT

\subsection{Supernova core collapse with microphysical equation of
  state and neutrino transport}
\label{subsec:sophisticated_core_collapse}

Next we explore the moderately relativistic regime of stellar core
collapse with the microphysical EoS of \citet{lattimer_91_a} and
neutrino transport. We simulated the collapse and the post-bounce
evolution of the progenitor model s15s7b2 with the \textsc{Vertex}
code as detailed in Sect.~\ref{subsec:transport_code}. The
calculations were performed using the TOV potential given in
Eq.~(\ref{eq:tov_potential}) \citep[Model V-R, which is identical with
the \textsc{Vertex} calculation of Model G15
in][]{liebendoerfer_05_a}, and we also tested the
modifications~A, B, and~F of the TOV potential (Models V-A, V-B, and
V-F, respectively). For comparison, we refer in the following
discussion also to the calculation of \citet{liebendoerfer_05_a} with
the fully relativistic \textsc{Agile-BoltzTran} code (Case~GR, Model
AB-GR).

\begin{figure}
  \resizebox{\hsize}{!}
  {\includegraphics{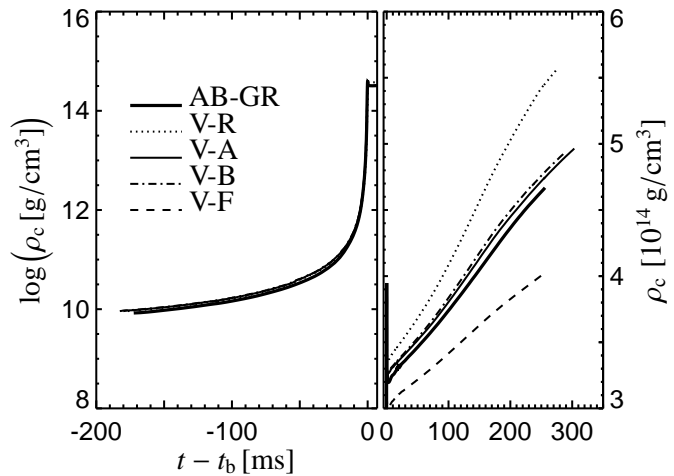}}
  \caption{Time evolution of the central density $ \rho_\mathrm{c} $
    for Model AB-GR (bold solid line), V-A (solid line), V-B
    (dashed-dotted line), V-F (dashed line), and V-R (dotted
    line). The left panel shows the collapse phase (note that here all
    models with the \textsc{Vertex} code lie on top of each other),
    while the right panel shows the post-bounce evolution. Note the
    different axis scales in both panels.}
  \label{fig:vertex_den}
\end{figure}

Fig.~\ref{fig:vertex_den} shows the central density as a function of
time for the collapse (left panel) and for the subsequent post-bounce
phase (right panel). The ``central'' density is the density value at
the centre of the innermost grid zone of the AB-GR simulation. Because
of a different numerical resolution it was necessary to interpolate
the \textsc{Vertex} results to this radial position. During the
collapse only minor differences between the relativistic calculation
(bold solid line) and the calculations with the \textsc{Vertex} code
are visible. Note that the trajectories from the \textsc{Vertex} code,
with the modifications~A, B, and~F of the TOV potential as well as
with the TOV potential (case~R), lie on top of each other.

We can thus infer that the differences between the modifications~A, B,
and~F of the TOV potential are unimportant during the collapse
phase. Furthermore, we can conclude from the good agreement of the
general relativistic calculation and the \textsc{Vertex} calculations
that the TOV potential works well during the collapse phase. However,
after core bounce this potential overestimates the compactness of the
forming neutron star (just as in the purely hydrodynamic simulations
in Sect.~\ref{subsec:simple_core_collapse}), and therefore the density
trajectories of Model AB-GR and Model V-R diverge (see
Fig.~\ref{fig:vertex_den}). At $ 250 \mathrm{\ ms} $ after the shock
formation the central density in Model V-R is about 20\% higher than
the one in the relativistic calculation. At this time the
modifications~A and~B of the TOV potential give a central density only
about 2\% higher than Model AB-GR, and the absolute difference stays
practically constant during the entire post-bounce evolution. This
implies that both modifications yield very good quantitative agreement
with the general relativistic treatment. In contrast, in Model V-F the
central density after bounce is lower than the relativistic result of
Model AB-GR. This indicates a strong underestimation of the depth of
the gravitational potential in Case~F, where $ \Gamma = 1 $ in the
integrand of Eq.~(\ref{eq:tov_potential}).

Since the central densities suggest that differences between a fully
relativistic calculation and Newtonian simulations with effective
relativistic potential become significant only after shock formation
\citep[see also][]{liebendoerfer_05_a}, we discuss the implications of
our potential modifications in the following only during the
post-bounce evolution.

\begin{figure}
  \resizebox{\hsize}{!}
  {\includegraphics{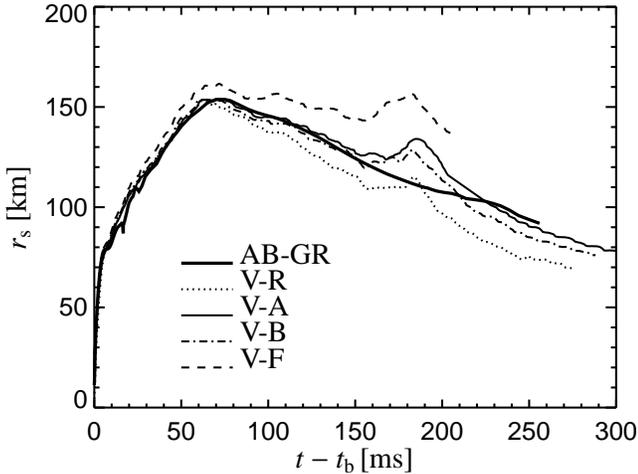}}
  \caption{Time evolution of the shock position $ r_\mathrm{s} $ for
    simulations with the \textsc{Vertex} code using various
    modifications of the TOV potential compared to the general
    relativistic result from the \textsc{Agile-BoltzTran} code (Model
    AB-GR, thick solid line).}
  \label{fig:vertex_shock}
\end{figure}

Fig.~\ref{fig:vertex_shock} shows the shock positions as functions of
time. Both Case~A (thin solid line) and Case~B (dashed-dotted
line) reveal the desirable trend of a closer match with the general
relativistic calculation (thick solid line) than seen for Model V-R,
which gives a shock radius that is too small, and Model V-F, where the
shock is too far out at all times. In particular, Model V-A reveals
excellent agreement with Model AB-GR. The only major difference is
visible between $ 170 \mathrm{\ ms} $ and about $ 230 \mathrm{\ ms} $
after shock formation when the shock transiently expands in the
\textsc{Vertex} calculation. This behaviour is generic for the
\textsc{Vertex} results and independent of the choice of the
gravitational potential. In the \textsc{Agile-BoltzTran} run the
transient shock expansion is much less pronounced and also a bit
delayed relative to the \textsc{Vertex} feature (it is visible as a
deceleration of the shock retraction between about
$ 200 \mathrm{\ ms} $ and $ 250 \mathrm{\ ms} $). This difference,
however, is not caused by general relativistic effects but is a
consequence of a different numerical tracking of the time evolution of
interfaces between composition layers in the collapsing stellar core
\citep[for more details about the numerics and a discussion of the
involved physics, see][]{liebendoerfer_05_a}. It is therefore
irrelevant for our present comparison of approximations to general
relativity. A good choice for the effective relativistic potential
(like Case~A) should just ensure that the corresponding shock
trajectory converges again with the relativistic result (Case~GR)
after the transient period of shock expansion.

The time evolution of the central density or the shock position,
however, is not the only relevant criterion for assessing the quality
of approximations to general relativity, as already discussed in
Sect.~\ref{subsec:simple_core_collapse}. A good approximation does not
only require good agreement for particular time-dependent quantities
(like, e.g., the central density), but also requires that the radial
structure of the models reproduces the relativistic case as well as
possible at any time.

\begin{figure*}[!t]
  \begin{tabular}{cc}
    \includegraphics[width=0.47\linewidth]{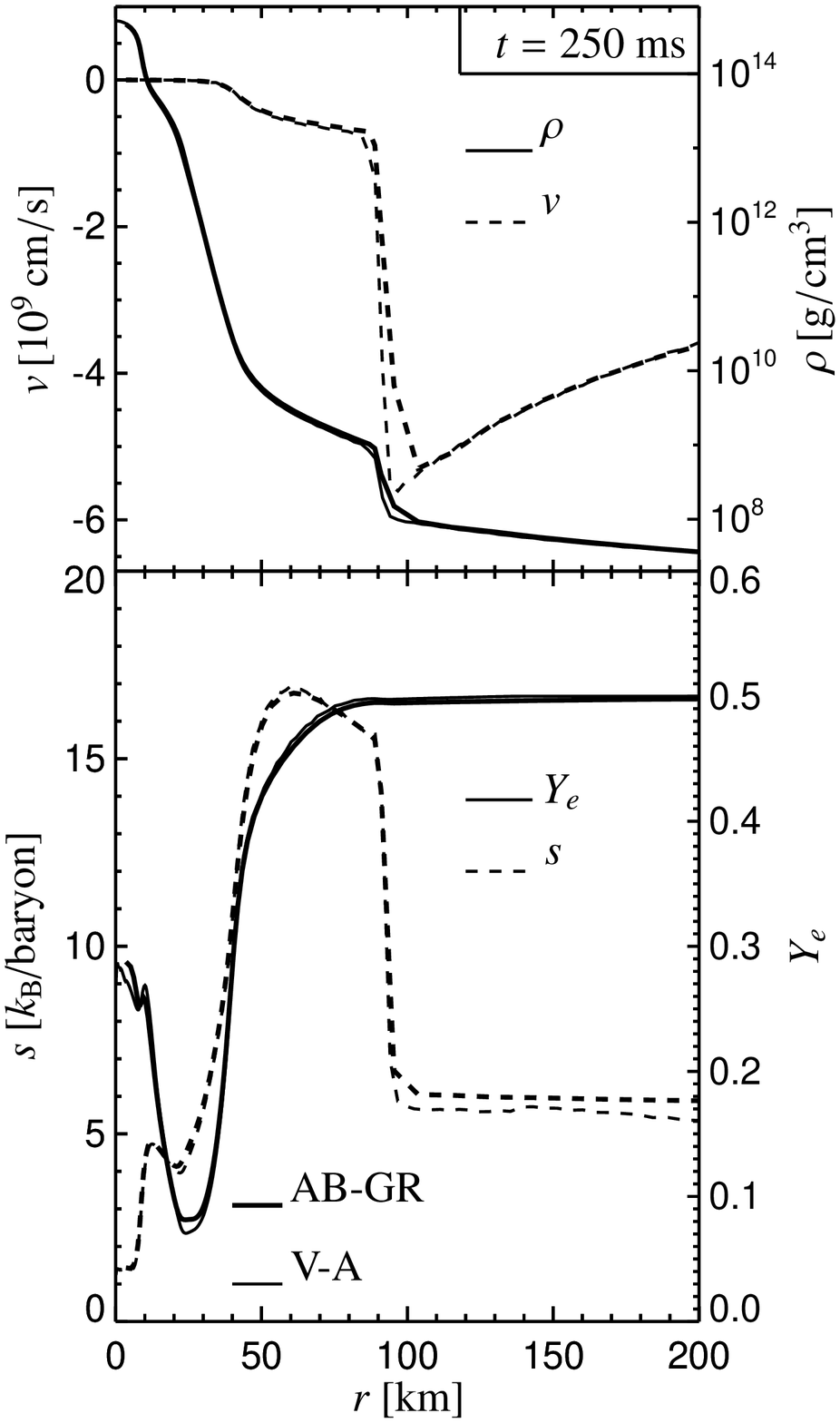} &
    \includegraphics[width=0.47\linewidth]{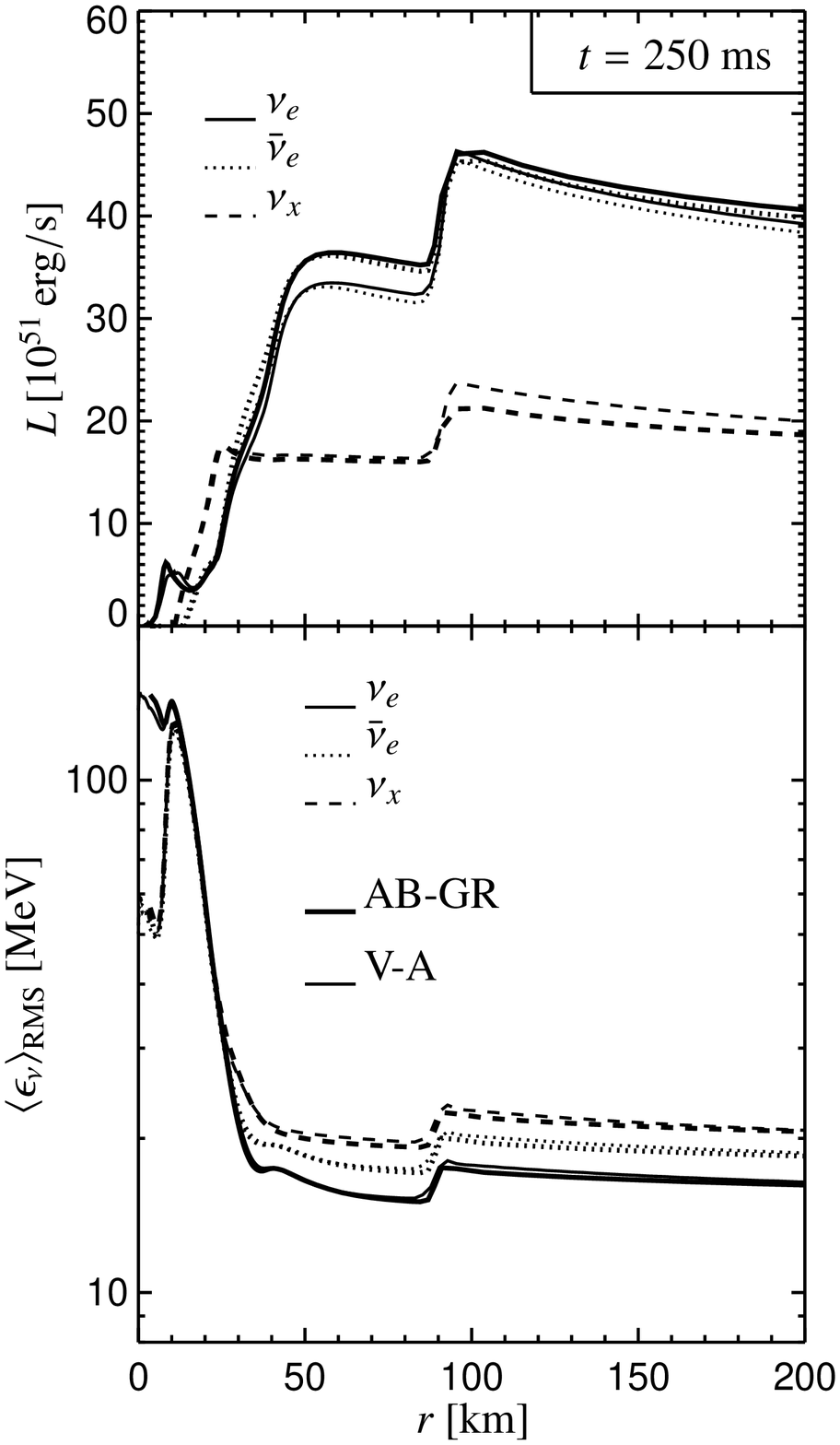}
  \end{tabular}
  \caption{Left: Radial profiles of the velocity $ v $ (dashed lines)
    and density $ \rho $ (solid lines) for Model V-A (thin) and the
    relativistic Model AB-GR (bold) at a time of $ 250 \mathrm{\ ms} $
    after shock formation (top panel), as well as radial profiles of
    the entropy $ s $ (dashed lines) and of the electron fraction
    $ Y_e $ (solid lines) for the same models and the same time
    (bottom panel). As in \citet{liebendoerfer_05_a}, $ r $ is the
    circumferential radius in case of the relativistic results.
    Note the different vertical axes on both sides of
    the two panels. Right: Radial profiles of the luminosities $ L $
    of electron neutrinos (solid lines), electron anti-neutrinos
    (dotted lines), and heavy-lepton neutrinos (dashed lines) for
    Models V-A (thin) and AB-GR (bold) at a time of
    $ 250 \mathrm{\ ms} $ after shock formation (top panel), as well
    as radial profiles of the root mean square energies
    $ \langle \epsilon_\nu \rangle_\mathrm{RMS} $ for the number
    densities of $ \nu_e $, $ \bar{\nu}_e $, and heavy-lepton
    neutrinos for Models V-A and AB-GR (bottom panel). The labeling is
    the same as in the panel above, and all neutrino quantities are
    given for a comoving observer.}
  \label{fig:vertex_profiles_nur_gr}
\end{figure*}

In the left panels of Fig.~\ref{fig:vertex_profiles_nur_gr} we show
such profiles of the density and velocity (top panel) and of the
entropy and electron fraction $ Y_e $, i.e., the electron-to-baryon
ratio (bottom panel), for Models AB-GR and V-A at a time of
$ 250 \mathrm{\ ms} $ after bounce, when the discrepancy between
relativistic and approximative treatment was found to be largest in
\citet{liebendoerfer_05_a}. Fig.~\ref{fig:vertex_profiles_nur_gr} can
be directly compared with Fig.~12 in the latter reference. Obviously
Model V-A fits the density profile of the general relativistic
calculation (Model AB-GR) extremely well at all radii. Furthermore,
both the velocity ahead of the shock front and behind it are in
extremely good agreement between the two models (the differences at
the shock jump have probably a numerical reason associated with
the different handling of shock discontinuities in both codes). It is
an important result that with this modification~A of the TOV potential
one is able to approximate the kinematics of the relativistic run with
astonishingly good quality in a Newtonian calculation (at least in
supernova simulations when the velocities do not become highly
relativistic). In contrast, in \textsc{Vertex} runs with the original
TOV potential (Case~R) the pre-shock velocities were found to be
significantly too large \citep{liebendoerfer_05_a}, which is due to
the overestimated strength of gravity in the far field limit as
compared to the relativistic calculations (see the discussion in
Sect.~\ref{subsec:theoretical_motivation}).

Also the entropy and $ Y_e $ profiles (bottom left panel of
Fig.~\ref{fig:vertex_profiles_nur_gr}) reveal a similarly excellent
agreement between Models V-A and AB-GR. The minor entropy differences
ahead of the shock are associated with a slightly different
description of the microphysics (nuclear burning and equation of
state) in the infall region \citep[for details we refer
to][]{liebendoerfer_05_a}.

Not only the radial structure of the forming neutron star in all
relevant quantities is well reproduced, but also the neutrino
transport results of the relativistic calculation and of the
approximative description of Case~A are in nearly perfect
agreement. Corresponding radial profiles of the luminosities and root
mean square energies -- both as defined in Sect.~4 of
\citet{liebendoerfer_05_a} -- for electron neutrinos, $ \nu_e $,
electron antineutrinos, $ \bar{\nu}_e $, and heavy-lepton
neutrinos\footnote{Since the transport of muon and tau neutrinos and
  antineutrinos differs only in minor details we treat all
  heavy-lepton neutrinos identically in the \textsc{Vertex}
  simulations.}, $ \nu_x $, are displayed in the right panels of
Fig.~\ref{fig:vertex_profiles_nur_gr}. The results for Models V-A and
AB-GR for all neutrino flavours share their characteristic features,
and in particular agree in the radial positions where the different
luminosities start to rise. While the $ \nu_x $ luminosities are
nearly indistinguishable below the shock, the jump at the shock
is slightly higher for the \textsc{Vertex} run and reflects the
larger effects due to observer motion, e.g., Doppler blueshift and
angular aberration, for an observer comoving with the rapidly
infalling stellar fluid ahead of the shock. The offset between results
of Models~V-A and AB-GR decreases at larger radii where the infall
velocities are lower. This discrepancy was not discovered by
\citet{liebendoerfer_05_a}, because there the agreement of the radial
structure for both investigated models was generally found to be
poorer than in the present work.

\begin{figure}
  \resizebox{\hsize}{!}
  {\includegraphics{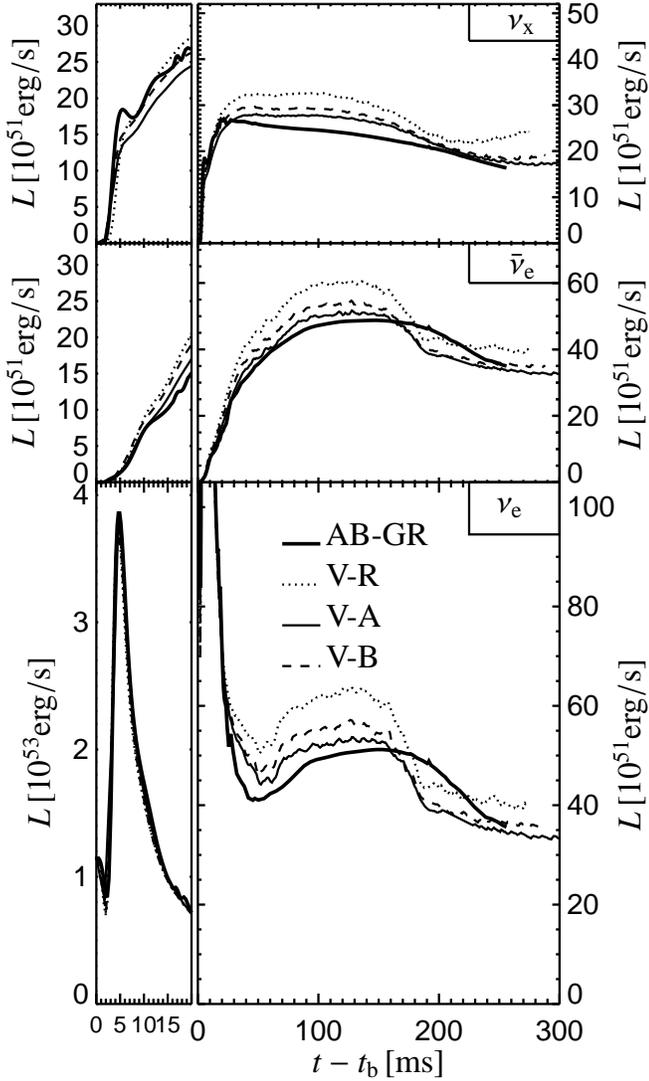}}
  \caption{Luminosities as functions of post-bounce time for different
    cases computed with the \textsc{Vertex} code and for Model
    AB-GR. The top panel shows the results for heavy-lepton neutrinos,
    the centre panel those for the electron antineutrinos, and the
    bottom panel the results for electron neutrinos. The panels on the
    left magnify the early post-bounce phase. All luminosities are
    given for an observer comoving with the stellar fluid at a radius
    of $ 500 \mathrm{\ km} $. Note the different scales of the
    vertical axes.}
  \label{fig:vertex_burst}
\end{figure}

\begin{figure}
  \resizebox{\hsize}{!}
  {\includegraphics{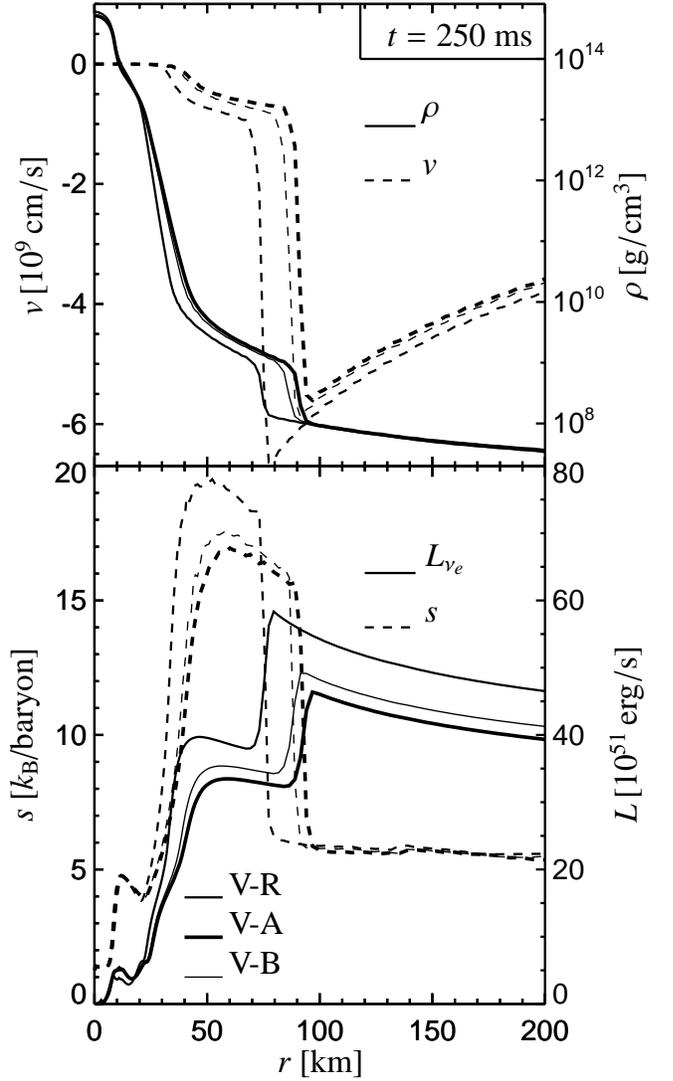}}
  \caption{Radial profiles of the velocity $ v $ (dashed lines) and
    density $ \rho $ (solid lines) for Models V-A (bold lines), V-B
    (thin), and V-R (medium) at a time of $ 250 \mathrm{\ ms} $ after
    shock formation (top panel). Radial profiles of the entropy $ s $
    (dashed lines) and of the electron neutrino luminosity
    $ L_{\nu_e} $(solid lines) for the same models and the same time
    (bottom panel). The luminosity is given for an observer comoving
    with the stellar fluid.}
  \label{fig:vertex_entropy_luminosity_profile}
\end{figure}

General relativistic effects are unlikely as an explanation,
because they are very small around the shock \citep[see Fig.~13
in][]{liebendoerfer_05_a}. A detailed analysis reveals that both codes
produce internally consistent results, conserving to good
precision the luminosity through the shock for an observer
at rest and showing the expected and physically correct behavior
in the limit of large radii. Most of the observed difference 
(which has no mentionable significance for supernova modeling) could
be traced back to the fact that \textsc{Vertex} achieves only order
$ (v / c) $ accuracy, whereas \textsc{Agile-BoltzTran} produces the
full relativistic result including higher orders in $ (v / c) $.
Corresponding effects become noticeable when $ v / c \ga 0.1 $. The
mean neutrino energies are hardly affected by this difference
(Fig.~\ref{fig:vertex_profiles_nur_gr}, bottom right panel).
In case of the $ \nu_e $ and $ \bar{\nu}_e $ luminosities the 
\textsc{Vertex} run yields roughly 10\% lower values outside
of the corresponding neutrino spheres (i.e., between about
$ 50 \mathrm{\ km} $ and $ 90 \mathrm{\ km} $), but values much closer
to those from the \textsc{Agile-BoltzTran} calculation ahead of the
shock. Since the neutrinospheric emission of $ \nu_e $ and
$ \bar{\nu}_e $ is strongly affected by the mass accretion rate of the
nascent neutron star and the corresponding accretion luminosity (which
both seem to have the tendency of being slightly higher in Model
AB-GR), we refrain from ascribing the different magnitude of the
$ \nu_e $ and $ \bar{\nu}_e $ luminosities only to the treatment of
relativistic effects. Although such a connection cannot be excluded,
the luminosity differences might (partly) also be a consequence of the
different accretion histories in Models AB-GR and V-A, which manifest
themselves in the shock trajectories (Fig.~\ref{fig:vertex_shock}) and
are attributable to the different handling of the microphysics and
computational grid in the infall layer \citep[see above
and][]{liebendoerfer_05_a}. This interpretation seems to be supported
by the time evolution of the neutrino luminosities plotted in
Fig.~\ref{fig:vertex_burst}. The accretion bump in the $ \nu_e $ and
$ \bar{\nu}_e $ luminosities which follows after the prompt $ \nu_e $
burst is stretched in time in case of Model AB-GR, indicating the
delay of mass infall at higher rates relative to all \textsc{Vertex}
simulations. Note that the neutrino emission reacts with a time lag of
some $ 10 \mathrm{\ ms} $ (corresponding to the cooling timescale of
the accretion layer on the neutron star) to variations of the mass
accretion rate.

Moreover, Fig.~\ref{fig:vertex_burst} shows that our variations of the
effective relativistic potential in the \textsc{Vertex} models have
little influence on the prompt burst of $\nu_e$ at shock breakout. But
subsequently the overestimated compactness of the proto-neutron star
in Model V-R, which causes the faster contraction of the stalled shock
after maximal expansion (Fig.~\ref{fig:vertex_shock}), also leads to
higher neutrino luminosities during the accretion phase. Consistent
with the shock trajectories, Model V-A yields the closest match with
the general relativistic run of Model AB-GR also for the neutrino
luminosities. It is very satisfactory that the results (shock radius
$ r_\mathrm{s} $ as well as the neutrino luminosities $ L $) from both
simulations reveal convergence at later times when the period of
massive post-bounce accretion comes to an end.

In Fig.~\ref{fig:vertex_entropy_luminosity_profile} we present the
radial structure at $ 250 \mathrm{\ ms} $ after bounce for the
\textsc{Vertex} simulations with the modifications~A and~B of the TOV
potential, compared to the results with the TOV potential (Case~R)
which was already discussed in \citet{liebendoerfer_05_a}. Note that
because of the excellent agreement seen in
Fig.~\ref{fig:vertex_profiles_nur_gr}, Model V-A (Case~A) can also be
considered as a representation of the fully relativistic run of Model
AB-GR. Models V-A and V-B show results of similar quality. The little
offset of the shock position (which is causally linked to the
differences in all profiles) might suggest that Model V-B is slightly
inferior to Model V-A in approximating relativity. This conclusion
could also be drawn from the post-bounce luminosities in
Fig.~\ref{fig:vertex_burst}. However, caution seems to be advisable
with such an interpretation, being aware of the uncertainties in the
accretion phase and infall layer discussed above, and in view of the
fact that the central densities (Fig.~\ref{fig:vertex_den}) and radial
density profiles (Fig.~\ref{fig:vertex_entropy_luminosity_profile})
agree well. Moreover, the quality of the agreement at ``very late''
times cannot be judged, because no information is available for the
behaviour of Model AB-GR after $ 250 \mathrm{\ ms} $ post bounce, a
time when the settling of the shock radius and luminosities to their
post-accretion levels seems not yet over in this model
(Figs.~\ref{fig:vertex_shock}, \ref{fig:vertex_burst}). The TOV
potential of Case~R clearly produces too large infall velocities ahead
of the shock (and therefore does not agree well with the kinematics of
the relativistic calculation), overestimates the compactness of the
forming neutron star, and thus produces too high neutrino luminosities
during the simulated period of evolution \citep[for a detailed
discussion, see][]{liebendoerfer_05_a}. Cases~A and~B clearly perform
better and must be considered as significant improvements for use in
Newtonian simulations with an effective relativistic potential as
approximations to fully relativistic calculations.

% END SUBSECTION SIMULATIONS RESULTS 1D NEUTRINO TRANSPORT
%%%%%%%%%%%%%%%%%%%%%%%%%%%%%%%%%%%%%%%%%%%%%%%%%%%%%%%%%%%%%%%%%
%%%%%%%%%%%%%%%%%%%%%%%%%%%%%%%%%%%%%%%%%%%%%%%%%%%%%%%%%%%%%%%%%

%%%%%%%%%%%%%%%%%%%%%%%%%%%%%%%%%%%%%%%%%%%%%%%%%%%%%%%%%%%%%%%%%
%%%%%%%%%%%%%%%%%%%%%%%%%%%%%%%%%%%%%%%%%%%%%%%%%%%%%%%%%%%%%%%%%
% SUBSECTION MIGRATION TEST

\subsection{Neutron stars in equilibrium -- Migration test}
\label{subsec:migration_test}

The supernova core collapse simulations presented in
Sect.~\ref{subsec:simple_core_collapse}
and~\ref{subsec:sophisticated_core_collapse} are limited to a maximum
core compactness $ 2 M_\mathrm{core} / R_\mathrm{core} \sim 0.2 $. In
order to extend the assessment of the quality of a Newtonian
simulation with an effective relativistic potential to a more strongly
relativistic regime, we construct spherical equilibrium models of
neutron stars using a simple polytropic EoS, $ P = K \rho^\gamma $,
with $ \gamma = 2 $ and $ K = 1.455 \times 10^5 $ (in cgs units). The
TOV equations are solved on a very fine grid with $ 10^5 $ equidistantly
spaced radial grid points using a second-order accurate Runge--Kutta
integration scheme. By varying the value of the central density from
$ \rho_\mathrm{c} = 1.0 \times 10^{14} \mathrm{\ g\ cm}^{-3} $ to
$ 100.0 \times 10^{14} \mathrm{\ g\ cm}^{-3} $, we cover the entire
range from weak to strong relativistic gravity. The density variation
corresponds to compactness parameters ranging from
$ 2 M_\mathrm{g} / R \sim 0.02 $ to $ 0.5 $, where $ M_\mathrm{g} $ is
the total gravitational mass of the neutron star and $ R $ its radius.
We again point out that in spherical symmetry the assumption of CFC
constitutes no approximation.

\begin{figure}
  \resizebox{\hsize}{!}
  {\includegraphics{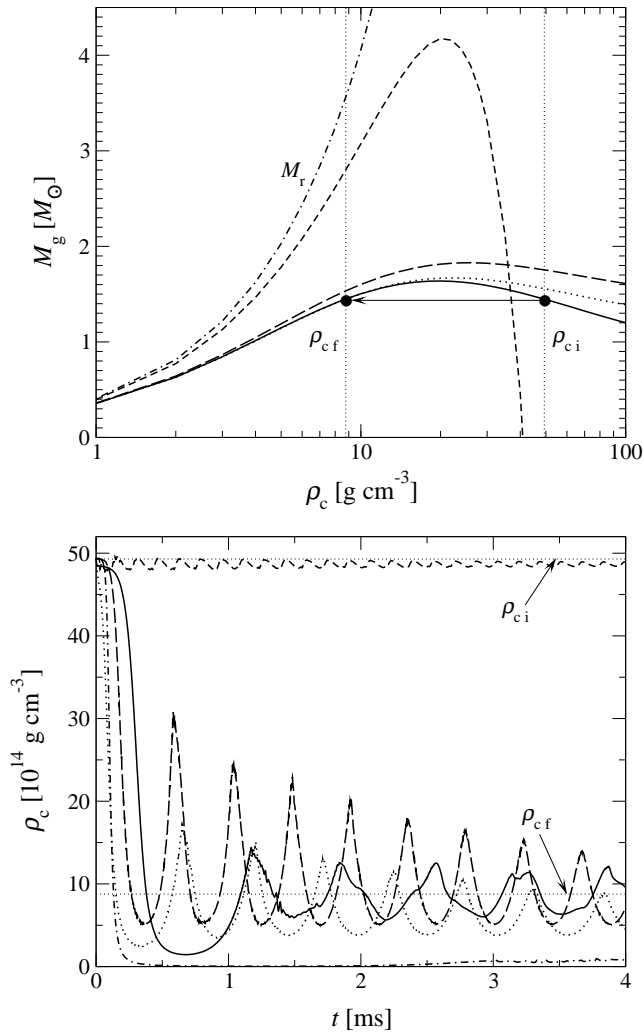}}
  \caption{Mass-density diagram for spherical equilibrium neutron star
    models (top panel) and time evolution of the central density for
    the migration test from the unstable to the stable branch (bottom
    panel). In both graphs the behaviour of the Newtonian models
    computed with the modification~A (dotted line) and~B (long-dashed
    line) of the TOV potential is close to that in the relativistic
    case (Case~GR; solid line). The Newtonian equivalent of the
    gravitational mass $ M_\mathrm{g} $ (dashed line) as well as the
    rest mass $ M_\mathrm{r} $ (dashed-dotted line) calculated using a
    regular Newtonian potential (Case~N) deviate strongly from the
    relativistic solution (top panel). While $ \rho_\mathrm{c} $
    oscillates around its initial value $ \rho_\mathrm{c\,i} $ for
    Case~N (dashed line), in Case~R (dashed-dotted line) the neutron
    star disperses to a state with
    $ \rho_\mathrm{c} \ll \rho_\mathrm{c\,f} $ (bottom panel).}
  \label{fig:migration_test}
\end{figure}

The mass-density curve of the corresponding TOV neutron star models
(Fig.~\ref{fig:migration_test}, top panel) exhibits the well-known
form consisting of a stable branch at low densities and an unstable
branch at high densities, where the gravitational mass
$ M_\mathrm{g} $ increases or decreases with $ \rho_\mathrm{c} $,
respectively. The two branches meet at the maximum
mass. Configurations on the unstable branch are unstable to small
perturbations and either collapse to a black hole or migrate to a
configuration on the stable branch with the same gravitational mass
but a lower central density
$ \rho_\mathrm{c\,f} < \rho_\mathrm{c\,i} $ (as indicated by the arrow
for an exemplary configuration in the top panel of
Fig.~\ref{fig:migration_test}).

While the gravitational mass $ M_\mathrm{g} $ and the rest mass
$ M_\mathrm{r} $ are limited in relativistic gravity (Case~GR),
$ M_\mathrm{g} \simeq M_\mathrm{r} \simeq 1.6 M_\odot $, regular
Newtonian gravity (Case~N) allows for solutions with arbitrarily high
rest mass $ M_\mathrm{r} $, which are all stable against (small)
perturbations. If an (unphysical) equivalent to the relativistic
gravitational mass is introduced in Newtonian gravity,
$ M_\mathrm{g} \equiv M_\mathrm{r} - M_\mathrm{b} $, where
$ M_\mathrm{b} \equiv |E_\mathrm{b}| $ is the mass associated with the
binding energy of the self-gravitating object, we find negative values
at large densities, where the binding energy becomes so large that
$ M_\mathrm{b} > M_\mathrm{r} $.

On the other hand, when using the modifications~A and~B of the TOV
potential in the TOV equation for the neutron star model sequence,
the corresponding curves for the TOV gravitational mass
$ M_\mathrm{g} $ in Fig.~\ref{fig:migration_test} agree very well with
the relativistic curve (Case~GR) even in the high density regime, and
they show the same qualitative behaviour with an increasing and a
decreasing branch. The respective mass maxima are located close to the
relativistic one for both modifications of the TOV potential.

Note that for constructing equilibrium configurations using the TOV
potential (Case~R) is identical to solving the fully relativistic TOV
equations (Case~GR), and thus the results are identical. Therefore,
this test is useless for predicting the quality of the TOV potential
in the context of a dynamic evolution. This is different when a
neutron star model from the unstable branch is allowed to evolve in
time using a dynamic code. Then the (small but nonzero) finite
difference discretisation errors of the numerical scheme act as
perturbations of the unstable equilibrium and excite small
oscillations. Depending on the numerical algorithms used in the code
and the resolution of the grid, the neutron star either contracts to
higher densities (and ultimately to a black hole for a simple
polytropic EoS) or approaches a new equilibrium state on the stable
branch after a strongly dynamical, nonlinear evolution phase. This
neutron star migration scenario is a standard test of the relativistic
regime with strongly dynamical evolution for numerical hydrodynamic
codes with relativistic gravity \citep[see, e.g.,][]{font_02_a,
  baiotti_04_a}.

The time evolution of the central density is depicted in the bottom
panel of Fig.~\ref{fig:migration_test} for such a migration to the
stable branch, where the central density $ \rho_\mathrm{c} $ drops from
$ \rho_\mathrm{c\,i} = 49.3 \times 10^{14} \mathrm{\ g\ cm}^{-3} $ to
$ \rho_\mathrm{c\,f} = 8.8 \times 10^{14} \mathrm{\ g\ cm}^{-3} $.
Shown is the evolution for a relativistic simulation (Case~GR) as
well as for Newtonian simulations utilizing a regular Newtonian
potential (Case~N), the TOV potential (Case~R), and its
modifications~A and~B. For these simulations we again use the
\textsc{CoCoNuT} code, whose implementation of the hybrid EoS,
Eqs.~(\ref{eq:hybrid_eos}, \ref{eq:hybrid_eos_terms}) allows one to
suppress the contribution of the thermal pressure $ P_\mathrm{th} $,
in which case the hybrid EoS reduces to a polytropic one.

As expected and as observed in simulations with similar relativistic
codes, in relativistic gravity (Case~GR) the neutron star initially
expands rapidly and its central density decreases strongly. It
undershoots the stable equilibrium at $ \rho_\mathrm{c\,f} $ and
experiences several ring-down oscillations of decreasing amplitude
until $ \rho_\mathrm{c} $ approaches $ \rho_\mathrm{c\,f} $ at late
times. The modifications~A and~B of the TOV potential both have initial
(final) states on the unstable (stable) branch of the mass-density
curve which are very close to the relativistic one
(Case~GR). Therefore the evolution of $ \rho_\mathrm{c} $ (starting
again with the same central density $ \rho_\mathrm{c\,i} $) is similar
to the relativistic case, as shown in
Fig.~\ref{fig:migration_test}. The disagreement in oscillation
amplitudes and periods can be attributed to the differences in the
Newtonian and relativistic kinematics. When comparing the results of
migration tests using various relativistic hydrodynamic codes
\citep[][]{stergioulas_04_a}, we observe that even small variations in
the numerical approach, e.g., a different coordinate choice or grid
setup, or differences in the treatment of the artificial atmosphere,
can have a strong impact on the detailed evolution of
$ \rho_\mathrm{c} $ from the initial to the final equilibrium
state. Consequently, we consider the results obtained with the
modifications~A or~B of the TOV potential in an otherwise Newtonian
formulation to be a very good approximation of the relativistic one
(Case~GR).

Using regular Newtonian gravity (Case~N), however, yields a totally
different behaviour, because there exist no unstable configurations in
that case. Thus, discretisation errors cannot trigger a migration, and
the star simply oscillates around its initial equilibrium state with
$ \rho_\mathrm{c} (t) \simeq \rho_\mathrm{c\,i} $. On the other hand,
using the TOV potential (Case~R) leads to a rapid dispersion (without
ring-down oscillations) of the neutron star with a final central
density $ \rho_\mathrm{c} \ll \rho_\mathrm{c\,f} $, which is in strong
qualitative disagreement with the behaviour in relativistic gravity
(Case~GR). We thus conclude that in the demanding test case of a
migrating spherical neutron star model, both the regular Newtonian
potential (Case~N) and the TOV potential (Case~R) fail, while the
modifications~A and~B of the TOV potential reproduce the relativistic
result very well, both qualitatively and also quantitatively.

% END SUBSECTION MIGRATION TEST
%%%%%%%%%%%%%%%%%%%%%%%%%%%%%%%%%%%%%%%%%%%%%%%%%%%%%%%%%%%%%%%%%
%%%%%%%%%%%%%%%%%%%%%%%%%%%%%%%%%%%%%%%%%%%%%%%%%%%%%%%%%%%%%%%%%

% END SECTION NUMERICAL RESULTS 1D
%%%%%%%%%%%%%%%%%%%%%%%%%%%%%%%%%%%%%%%%%%%%%%%%%%%%%%%%%%%%%%%%%
%%%%%%%%%%%%%%%%%%%%%%%%%%%%%%%%%%%%%%%%%%%%%%%%%%%%%%%%%%%%%%%%%

%%%%%%%%%%%%%%%%%%%%%%%%%%%%%%%%%%%%%%%%%%%%%%%%%%%%%%%%%%%%%%%%%
%%%%%%%%%%%%%%%%%%%%%%%%%%%%%%%%%%%%%%%%%%%%%%%%%%%%%%%%%%%%%%%%%
% SECTION NUMERICAL RESULTS 2D

\section{Multi-dimensional simulations with rotation}
\label{sec:rotating_simulations}

For the multi-dimensional calculations we only use the
\textsc{CoCoNuT} code, since parameter studies with \textsc{MudBath}
(the two-dimensional version of \textsc{Vertex}) are computationally
too expensive when neutrino transport is included. Moreover, currently
there exists no other fully relativistic two-dimensional code
including neutrino transport to compare our results with. The
following discussion hence focuses on purely hydrodynamic simulations
using the simple matter model described in
Sect.~\ref{subsec:hydro_code}.

\begin{figure}[ht!]
  \resizebox{\hsize}{!}
  {\includegraphics{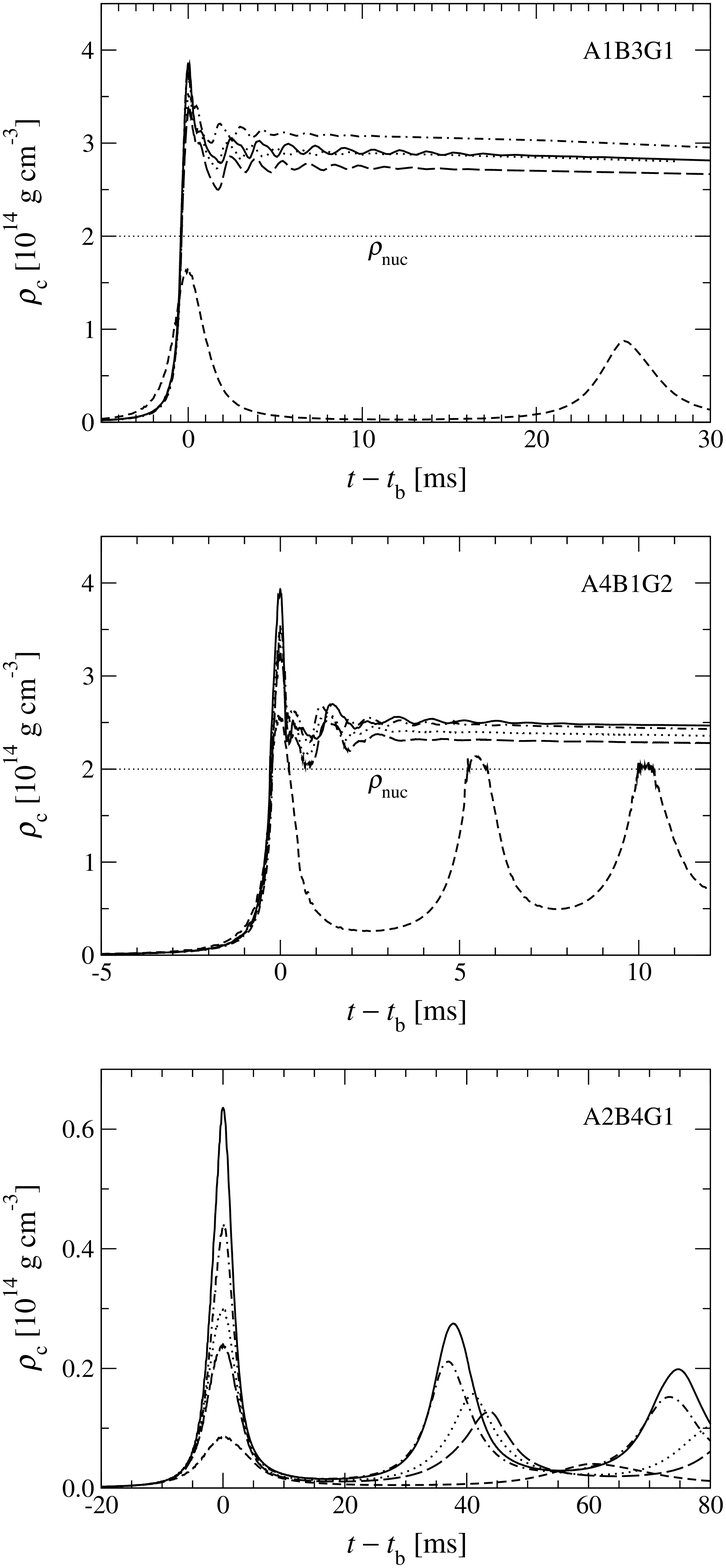}}
  \caption{Time evolution of the central density $ \rho_\mathrm{c} $
    for the rotating core collapse models A1B3G1 (top panel), A4B1G2
    (centre panel), and A2B4G1 (bottom panel), respectively. With
    increasing influence of rotation on the collapse dynamics, the
    results of the Newtonian simulations with the modifications~A
    (dotted line) and~B (long-dashed line) of the TOV potential and
    eventually also with the TOV potential (Case~R; dashed-dotted
    line) move away from the relativistic result (Case~GR; solid
    line). The central density obtained with regular Newtonian gravity
    (Case~N; dashed line) is lowest for all three models.}
  \label{fig:rotation_collapse}
\end{figure}

Motivated by the promising results in spherical symmetry, we now apply
the modified effective relativistic potentials according to
Eq.~(\ref{eq:effective_potential_multid}) to simulations of
rotational core collapse in axisymmetry. The initial models of the
pre-collapse core are described by the same EoS parameters as the ones
in Sect.~\ref{subsec:simple_core_collapse} with
$ \rho_{\rm c\,i} = 10^{10} {\rm\ g\ cm}^{-3} $, but now rotate at
various rotation rates with different degrees of differential
rotation. To construct these rotating equilibrium configurations, we
use Hachisu's self-consistent field method
\citep[see][]{komatsu_89_a}. The initial pressure reduction to trigger
the collapse and the stiffening of the EoS follows the prescription
given in Sect.~\ref{subsec:simple_core_collapse}.

For these multi-dimensional tests we choose three rotational core
collapse models as representative cases. Their evolution is influenced
by the effects of rotation at various degrees \citep[for details about
the specifications of these models, see][]{dimmelmeier_02_a}. The core
of Model A1B3G1 is an almost uniform rotator with a moderate rotation
rate and collapses nearly spherically, forming a compact remnant
immediately after core bounce. Model A4B1G2 is a very differentially
rotating model, which develops considerable deviations from spherical
symmetry during its evolution, but otherwise shares the qualitative
bounce behavior of the previous model A1B3G1. Finally, in
the multiple bounce model A2B4G1 the increase of the centrifugal
forces during the contraction phase causes a series of subsequent
multiple bounces at subnuclear densities. The time evolution of the
central density for these three models computed relativistically
(Case~GR) is shown with solid lines in the top, centre, and bottom
panels of Fig.~\ref{fig:rotation_collapse}, respectively.

In a Newtonian simulation with regular Newtonian gravity (Case~N) all
three models exhibit several consecutive bounces (at subnuclear
densities for Models A1B3G1 and A2B4G1, and just above nuclear matter
densities for Model A4B1G2), with the central density being considerably
lower during the entire evolution (Fig.~\ref{fig:rotation_collapse}).
Such a qualitative change of the collapse dynamics between a Newtonian
and a general relativistic simulation of the same model was also
observed in the comprehensive parameter study of rotational core
collapse performed by \citet{dimmelmeier_02_b}.

When applying the TOV potential (Case~R) or its modifications~A and~B
to Model A1B3G1 (top panel of Fig.~\ref{fig:rotation_collapse}) the
outcome is similar to the situation in spherical symmetry shown in
Fig.~\ref{fig:comparison_density_evolution}. Case~R yields a central
density noticeably higher than the one in the relativistic simulation
(Case~GR). For Case~A the result agrees very well with the
relativistic one, while for Case~B the central density is too small.

As the influence of rotation increases, the ordering of the curves for
Cases~R, A, and~B (from higher to lower $ \rho_\mathrm{c} $)
remains unaltered. However, all curves are shifted to lower densities
(Fig.~\ref{fig:rotation_collapse}). For Model A4B1G2 the result from
the Newtonian simulation with the TOV potential (Case~R) matches the
relativistic result (Case~GR) best, which is a coincidence associated
with the particular rotational state of this model. For Model A2B4G1,
where the influence of rotation is stronger, the results obtained for
Cases~A, B, and~R differ considerably from the relativistic results.

A behaviour consistent with that of these three representative models
is observed for other rotational core collapse models as well. Thus,
in rotational core collapse, even a Newtonian simulation with the
TOV potential (Case~R) yields lower central densities than the
corresponding relativistic simulation (Case~GR) beyond a certain
rotation rate (which depends both on the initial model and on the EoS
parameters).

We therefore conclude that for rotational core collapse the use of
neither the modifications~A and~B of the TOV potential nor the
(stronger) TOV potential itself (Case~R) can closely
reproduce the collapse behaviour of a relativistic code for all degrees
of rotation. Hence, in a Newtonian code either the effective
relativistic potential has to include the effect of rotation, or the
collapse kinematics (and thus the hydrodynamic equations) have to be
modified. However, in the latter case the advantage of a Newtonian
approach over a fully relativistic formulation in terms of simplicity
and computational speed may be lost.

On the other hand, we find that the qualitative dynamical behaviour of
relativistic rotational core collapse models (Case~GR), particularly
their collapse type \citep[regular collapse, multiple bounce collapse,
or rapid collapse as introduced by][]{zwerger_97_a} is correctly
captured in a Newtonian simulation for a wide range of parameter
space, if the TOV potential (Case~R) or its modified versions (Cases~A
to~D) are used. Moreover, the corresponding results agree much better
with the relativistic result than that obtained from a Newtonian
simulation with regular Newtonian potential (Case~N). Hence, the usage
of the effective relativistic potentials is an improvement of
Newtonian simulations also in the case of rotational core collapse.

% END SECTION NUMERICAL RESULTS 2D
%%%%%%%%%%%%%%%%%%%%%%%%%%%%%%%%%%%%%%%%%%%%%%%%%%%%%%%%%%%%%%%%%
%%%%%%%%%%%%%%%%%%%%%%%%%%%%%%%%%%%%%%%%%%%%%%%%%%%%%%%%%%%%%%%%%

%%%%%%%%%%%%%%%%%%%%%%%%%%%%%%%%%%%%%%%%%%%%%%%%%%%%%%%%%%%%%%%%%
%%%%%%%%%%%%%%%%%%%%%%%%%%%%%%%%%%%%%%%%%%%%%%%%%%%%%%%%%%%%%%%%%
%SECTION SUMMARY

\section{Summary and Conclusions}
\label{sec:summary}

We investigated different modifications of the TOV potential used
as effective relativistic potentials in Newtonian hydrodynamics with
the aim to test their quality of approximating relativistic effects in
stellar core collapse and supernova simulations. This work was
motivated by a recent comparison of the neutrino
radiation-hydrodynamics codes \textsc{Agile-BoltzTran} and
\textsc{Vertex} used by the Oak Ridge-Basel collaboration and by the
Garching group, respectively \citep{liebendoerfer_05_a}. While the
former code is a fully relativistic implementation of neutrino
transport and hydrodynamics in spherical symmetry, the latter code
employs an approximative relativistic description by treating the
self-gravity of the stellar fluid with an effective relativistic
potential tailored from a comparison of the Newtonian and relativistic
equations of motion. Otherwise \textsc{Vertex} solves the Newtonian
equations of hydrodynamics, ignoring the effects of relativistic
kinematics and space-time curvature \citep{rampp_02_a}. The comparison
by \citet{liebendoerfer_05_a} shows very good agreement of the results
from both codes for the phases of stellar core collapse, bounce and
prompt shock propagation. However, it also revealed a moderate but
gradually increasing overestimation of the compactness of the nascent
neutron star during the subsequent accretion phase, accompanied by an
overestimation of the luminosities and mean energies of the emitted
neutrinos.

These systematic differences suggested that an improvement might be
possible by changing the effective relativistic potential such that
the strength of the gravitational field is reduced relative to the TOV
potential. To this end we considered a variety of modifications
(Cases~A--F; see Sect.~\ref{subsec:improved_potentials}) of the TOV
potential (which defines our reference Case~R), and compared
corresponding hydrodynamic as well as hydrostatic solutions with
regular Newtonian (Case~N) and fully relativistic calculations
(Case~GR). While some modifications did not provide any significant
advantage over the TOV potential, several options for the potential
were identified which allow for a much better reproduction of
relativistic results with Newtonian hydrodynamics. In particular,
reducing the TOV mass, Eq.~(\ref{eq:tov_mass}), that appears in the
expression of the TOV potential, turned out to be very successful in
this respect.

For example, in Case~A we introduced an additional metric factor
$ \Gamma < 1 $, Eq.~(\ref{eq:gamma_factor}), in the integrand of the
TOV mass, and in Case~B we ignored the internal energy (and all terms
depending on neutrinos) in the TOV mass. In both cases not only
characteristic model parameters like the central density and shock
radius were in excellent agreement with relativistic simulations
during all stages of the evolution including the long-term accretion
phase of the collapsing core after bounce. Also the radial profiles
of the hydrostatic central core as obtained in the hydrodynamic
simulations were found to be very good approximations of the profiles
from the fully relativistic treatment. Consequently, they are nearly
consistent with solutions of the TOV equations for equal central
density. Migration tests of extremely relativistic, ultradense,
hydrostatic configurations from the unstable to the stable branch of
the mass-density relation confirmed the good qualitative and
quantitative performance of the approximation even for such a
demanding test problem with a strongly dynamic transition from the
initial to the final state. This suggests that the use of a modified
TOV potential in an otherwise Newtonian hydrodynamics code also
ensures a good reproduction of relativistic kinematics.

In addition to these purely hydrodynamic simulations with the
\textsc{CoCoNuT} code, we performed runs including neutrino transport
and compared them to relativistic calculations with the
\textsc{Agile-BoltzTran} code of the Oak Ridge-Basel collaboration
\citep[cf.\ the simulations for Model G15
in][]{liebendoerfer_05_a}. Also in this application, the
modifications~A and~B of the TOV potential produced a major
improvement relative to the original TOV potential (Case~R), for which
results with the \textsc{Vertex} code were already presented in
\citet{liebendoerfer_05_a}. Good agreement between the relativistic
and approximative treatments was seen in the transport quantities as
well. The \textsc{Vertex} code performs well despite the fact that the
approximative description takes into account only relativistic
redshift in the transport but -- for reasons of consistency with the
Newtonian hydrodynamics code -- ignores the difference between proper
radius and coordinate radius. Remaining (minor) quantitative
differences (below about 10\% for neutrino data and even smaller for
hydrodynamic quantities) are probably only partly associated with
relativistic effects. We suspect that a significant contribution to
these differences originates from small (but unavoidable because linked
to the specifics of the employed numerical methods) discrepancies in the
infall layer ahead of the shock as described by the \textsc{Vertex}
and \textsc{Agile-BoltzTran} codes.

The results discussed in this paper therefore demonstrate that a very
good, simple, and computationally efficient approximation to a
relativistic treatment of stellar core collapse and neutron star
formation in spherical symmetry can be achieved by modeling the
self-gravity of the stellar plasma with an effective relativistic
potential based on the TOV potential as suggested in this
work. Two-dimensional axisymmetric collapse calculations (using a
simple equation of state and ignoring neutrino transport) for rotating
stellar cores with a wide range of conditions showed that even in this
case the TOV potential and its modifications reproduce the
characteristics of the relativistic collapse dynamics quantitatively
well for not too rapid rotation. There is still good qualitative
agreement when the rotation becomes fast, in contrast to Newtonian
simulations with regular Newtonian potential (Case~N), which mostly
fail even qualitatively. Multi-dimensional simulations with an
effective relativistic potential (which in general does not satisfy
the Poisson equation), however, fulfill strict momentum conservation
only for special cases of symmetry (for details, see
Sect.~\ref{subsec:code_implementation}).

Although our Cases~A and~B yield results of similar quality for all the
test problems considered here and neither is preferred when a simple
equation of state is used, the situation is more in favour of Case~A
when microphysics is taken into account in the equation of state. When
particles can annihilate or nuclear interactions take place, rest mass
energy is converted into internal energy and vice versa. In such a
situation baryon number and lepton number are conserved, but not the
total rest mass of the particles, and only the total (``relativistic'')
energy density, $ \rho + e $, is a well defined quantity, but not the
individual energies. Case~B therefore becomes ambiguous
(see Sect.~\ref{subsec:improved_potentials}) while Case~A
does not suffer from such problems.
We therefore are tempted to recommend using Case~A as an effective
relativistic potential in Newtonian hydrodynamics codes.
The extra factor $ \Gamma < 1 $ in the
integrand of the TOV mass, Eq.~(\ref{eq:tov_mass}), reduces the mass
integral below the mass equivalent of the total (rest mass plus
internal) energy. Its introduction may be justified by
heuristic arguments and
consistency considerations when the relativistic concept of the
gravitating mass is applied in a Newtonian description of the
dynamics, which does not distinguish between proper volumes and
coordinate volumes (see Sect.~\ref{subsec:theoretical_motivation}).

%END SECTION SUMMARY
%%%%%%%%%%%%%%%%%%%%%%%%%%%%%%%%%%%%%%%%%%%%%%%%%%%%%%%%%%%%%%%%%
%%%%%%%%%%%%%%%%%%%%%%%%%%%%%%%%%%%%%%%%%%%%%%%%%%%%%%%%%%%%%%%%%

%%%%%%%%%%%%%%%%%%%%%%%%%%%%%%%%%%%%%%%%%%%%%%%%%%%%%%%%%%%%%%%%%
%%%%%%%%%%%%%%%%%%%%%%%%%%%%%%%%%%%%%%%%%%%%%%%%%%%%%%%%%%%%%%%%%
%SECTION ACKNOWLEDGMENTS

\begin{acknowledgements}
  We thank M.-A.~Aloy and M.~Ober\-gau\-linger for helpful discussions and
  M.~Rampp for his support. We are particularly grateful to the
  referee, M.~Liebend\"orfer, for his interest in our work and his
  knowledgeable comments, which helped us to improve our
  manuscript. We also thank him for providing us with data of his
  general relativistic run. This work was supported by the German
  Research Foundation DFG (SFB/Transregio~7
  ``Gravitations\-wellen\-astro\-nomie'' and SFB~375
  ``Astro\-teilchen\-physik''). The simulations were performed at the
  Max-Planck-Institut f\"ur Astrophysik and the RZG Rechenzentrum in
  Garching.
\end{acknowledgements}

%END SECTION ACKNOWLEDGEMENTS
%%%%%%%%%%%%%%%%%%%%%%%%%%%%%%%%%%%%%%%%%%%%%%%%%%%%%%%%%%%%%%%%%
%%%%%%%%%%%%%%%%%%%%%%%%%%%%%%%%%%%%%%%%%%%%%%%%%%%%%%%%%%%%%%%%%

\appendix

%%%%%%%%%%%%%%%%%%%%%%%%%%%%%%%%%%%%%%%%%%%%%%%%%%%%%%%%%%%%%%%%%
%%%%%%%%%%%%%%%%%%%%%%%%%%%%%%%%%%%%%%%%%%%%%%%%%%%%%%%%%%%%%%%%%
% SECTION CONSTRUCTION OF EOS (APPENDIX)

\section{Construction of an ``effective'' one-parameter equation of state}
\label{app:tov_solution}

The relativistic TOV solution for a spherical, self-gravitating matter
distribution in equilibrium can be found by solving the following two
coupled ordinary differential equations for the pressure $ P $
and the mass $ m $:
\begin{eqnarray}
  \frac{\partial P}{\partial r} & = &
  - \frac{\rho (1 + \epsilon) + P}{r^2}
  \left( m + 4 \pi r^3 P \right)
  \left( 1 - \frac{2 m}{r} \right)^{-1}\!\!\!\!\!\!\!,
  \label{eq:tov_equation_pressure}
  \\
  \frac{\partial m}{\partial r} & = &
  4 \pi \rho (1 + \epsilon) r^2.
  \label{eq:tov_equation_mass}
\end{eqnarray}%
with appropriate boundary conditions at the inner and outer boundary
at $ r = 0 $ and $ r = R $, respectively. Note that for a TOV solution
the mass $ m (R) $ is equivalent to the total gravitational mass
$ M_\mathrm{g} $ of the matter configuration.

The above system is closed by choosing a two-parameter EoS
$ P (\rho, \epsilon) $ and assuming a distribution profile for the
specific internal energy $ \epsilon $, which effectively transforms
the EoS into a one-parameter relation, $ P = P (\rho) $ [or
alternatively $ \rho = \rho (P) $], from which the specific internal
energy can be determined as $ \epsilon = \epsilon (P, \rho) $. A
typical example for this is to demand a polytropic relation for $ P $
and to specify the internal energy $ \epsilon $ by the ideal gas EoS:
\begin{equation}
  \begin{array}{l@{\quad}c@{\quad}l}
    P = K \rho^\gamma, & &
    \displaystyle \rho =
    \left( \frac{P}{K} \right)^{1 / \gamma}\!\!\!\!\!\!\!\!,
    \\ [-0.3 em]
    & \longrightarrow &
    \\ [-0.3 em]
    P = \rho \epsilon (\gamma - 1), & &
    \displaystyle \epsilon = \frac{P}{\rho (\gamma - 1)}.
  \end{array}
\end{equation}

For a two-parameter EoS of the form of Eqs.~(\ref{eq:hybrid_eos},
\ref{eq:hybrid_eos_terms}), where a priori no profile for the specific
internal energy is known, the following procedure can be applied to
construct an ``effective'' one-parameter EoS. For matter in equilibrium,
e.g., at late times long after the core bounce phase, both the radial
profiles of the pressure $ P (r) $ and the density $ \rho (r) $, as
well as the specific internal energy $ \epsilon (r) $ can be assumed
to be stationary. If the pressure profile is monotonic, then for each
value of $ P $ there exists a unique location $ r $. By inverting the
pressure profile, $ P (r) \longrightarrow r (P) $, we can thus relate
a specific value $ \rho (r) $ and $ \epsilon (r) $ to each value of
$ P $. This way a one-parameter EoS $ \rho = \rho (r (P)) $ [and
$ \epsilon = \epsilon (r (P)) $] can be constructed.

The excellent matching of the density profiles in the central part of
a collapsed core in equilibrium obtained from a dynamic evolution with
profiles from a TOV solution using equal central density and the EoS
transformed as described above (see
Figs.~\ref{fig:tov_comparison_density_profiles_1}
to~\ref{fig:tov_comparison_density_profiles_2}) demonstrates the
applicability of this method.

% END SECTION CONSTRUCTION OF EOS (APPENDIX)
%%%%%%%%%%%%%%%%%%%%%%%%%%%%%%%%%%%%%%%%%%%%%%%%%%%%%%%%%%%%%%%%%
%%%%%%%%%%%%%%%%%%%%%%%%%%%%%%%%%%%%%%%%%%%%%%%%%%%%%%%%%%%%%%%%%

%%%%%%%%%%%%%%%%%%%%%%%%%%%%%%%%%%%%%%%%%%%%%%%%%%%%%%%%%%%%%%%%%
%%%%%%%%%%%%%%%%%%%%%%%%%%%%%%%%%%%%%%%%%%%%%%%%%%%%%%%%%%%%%%%%%
% SECTION COORDINATE TRANSFORMATION (APPENDIX)

\section{Radial coordinate transformation}
\label{app:coordinate_transformation}

To compare a solution of the TOV
equations~(\ref{eq:tov_equation_pressure},
\ref{eq:tov_equation_mass}), which are formulated in standard
Schwarzschild-like coordinates, to results from our evolution code
using relativistic gravity, which is based on radial isotropic
coordinates, we must perform a coordinate transformation of the radial
coordinate.

In standard Schwarzschild-like coordinates, the line element
$ \mathrm{d}s_\mathrm{st}^2 $ reads
\begin{equation}
  \mathrm{d}s_\mathrm{st}^2 =
  - A \, \mathrm{d}t_\mathrm{st}^2 +
  B \, \mathrm{d}r_\mathrm{st}^2 +
  r_\mathrm{st}^2 \, \mathrm{d}\Omega_\mathrm{st}^2,
  \label{eq:schwarzschild_line_element}
\end{equation}
while the line element $ \mathrm{d}s_\mathrm{iso}^2 $ in radial
isotropic coordinates is given by
\begin{equation}
  \mathrm{d}s_\mathrm{iso}^2 =
  - \alpha^2 \, \mathrm{d}t_\mathrm{iso}^2 +
  \phi^4 \left( \mathrm{d}r_\mathrm{iso}^2 +
  r_\mathrm{iso}^2 \mathrm{d}\Omega_\mathrm{iso}^2 \right).
  \label{eq:isotropic_line_element}
\end{equation}
In vacuum outside the matter distribution, the metric components
become
\begin{equation}
  A = 1 - \frac{2 M_\mathrm{g}}{r_\mathrm{st}},
  \qquad
  B = \left( 1 - \frac{2 M_\mathrm{g}}{r_\mathrm{st}} \right)^{-1} \!\!\!\!\! =
  A^{-1}
  \label{eq:schwarzschild_vacuum_components}
\end{equation}
and
\begin{equation}
  \alpha = \frac{1 - \frac{M_\mathrm{g}}{2 r_\mathrm{iso}}}
  {1 + \frac{M_\mathrm{g}}{2 r_\mathrm{iso}}},
  \qquad
  \phi = 1 + \frac{M_\mathrm{g}}{2 r_\mathrm{iso}},
  \label{eq:isotropic_vacuum_components}
\end{equation}
respectively.

As with both metrics one can describe a spherical matter distribution
in equilibrium, equating the line elements,
$ ds_\mathrm{st}^2 = ds_\mathrm{iso}^2 $, yields conditions for each
of the coordinates and metric components in such a case.

Setting the two coordinate times equal, we obtain a condition for
$ A $ and $ \alpha $:
\begin{equation}
  \left.
  \begin{array}{rcl}
    A \, \mathrm{d}t_\mathrm{st}^2 & = &
    \alpha^2 \, \mathrm{d}t_\mathrm{iso}^2,
    \\ [0.3 em]
    \mathrm{d}t_\mathrm{st} & = &
    \mathrm{d}t_\mathrm{iso},
  \end{array}
  \right\}
  \quad \longrightarrow \quad
  \sqrt{A} = \alpha.
  \label{eq:time_transformation}
\end{equation}
A similar condition can be obtained for the angle element:
\begin{equation}
  \left.
  \begin{array}{rcl}
    r_\mathrm{st}^2 \, \mathrm{d}\Omega_\mathrm{st}^2 & = &
    \phi^4 r_\mathrm{iso}^2 \mathrm{d}\Omega_\mathrm{iso}^2,
    \\ [0.3 em]
    \mathrm{d}\Omega_\mathrm{st} & = & \mathrm{d}\Omega_\mathrm{iso},
  \end{array}
  \right\}
  \quad \longrightarrow \quad
  r_\mathrm{st} = \phi^2 r_\mathrm{iso}.
  \label{eq:angle_transformation}
\end{equation}
From the radial elements, we get a relation for the radial
differentials:
\begin{equation}
  \sqrt{B} \, \mathrm{d}r_\mathrm{st} =
  \phi^2 \, \mathrm{d}r_\mathrm{iso}.
  \label{eq:radius_transformation}
\end{equation}

Knowing the solution in the isotropic radial coordinate, it is
straightforward to derive a simple expression for
$ r_\mathrm{st} (r_\mathrm{iso}) $ from
Eq.~(\ref{eq:angle_transformation}).

On the other hand, obtaining the inverse relation
$ r_\mathrm{iso} (r_\mathrm{st}) $ is more complicated. For this, we
insert Eq.~(\ref{eq:angle_transformation}) into
Eq.~(\ref{eq:radius_transformation}) and arrive at
\begin{equation}
  \frac{\mathrm{d}r_\mathrm{iso}}{\mathrm{d}r_\mathrm{st}} =
  \frac{\sqrt{B}}{\phi^2} =
  \frac{r_\mathrm{iso}}{r_\mathrm{st}} \sqrt{B}.
  \label{eq:transformation_equation_1}
\end{equation}

This differential equation can be solved by numerical integration
using, e.g., a Runge--Kutta integration scheme with the following
boundary conditions:
\begin{eqnarray}
  r_\mathrm{0\,iso} & = & r_\mathrm{0\,st} = 0,
  \\
  R_\mathrm{iso} & = & \frac{R_\mathrm{st}}{2}
  \left( 1 - \frac{M_\mathrm{g}}{R_\mathrm{st}} +
  \sqrt{1 - \frac{2 M_\mathrm{g}}{R_\mathrm{st}}} \right).
  \label{eq:transformation_boundary_conditions}
\end{eqnarray}%

An alternative method is to rewrite
Eq.~(\ref{eq:transformation_equation_1}) in terms of
$ \ln r_\mathrm{iso} $,
\begin{equation}
  \frac{\mathrm{d}\ln r_\mathrm{iso}}{\mathrm{d}r_\mathrm{st}} =
  \frac{\sqrt{B}}{r_\mathrm{st}},
  \label{eq:transformation_equation_2}
\end{equation}
which can be directly integrated numerically.

% END SECTION COORDINATE TRANSFORMATION (APPENDIX)
%%%%%%%%%%%%%%%%%%%%%%%%%%%%%%%%%%%%%%%%%%%%%%%%%%%%%%%%%%%%%%%%%
%%%%%%%%%%%%%%%%%%%%%%%%%%%%%%%%%%%%%%%%%%%%%%%%%%%%%%%%%%%%%%%%%

\bibliography{paper}

\end{document}